\documentclass[fleqn,usenatbib]{mnras}
\pdfoutput=1
\usepackage{ae,aecompl}
\usepackage{graphicx}
\usepackage{media9}
\usepackage{amsmath}
\usepackage{amssymb}
\usepackage{enumerate}
\usepackage{txfonts}
\DeclareSymbolFont{cmletters}{OML}{cmm}{m}{it}
\DeclareMathSymbol{v}{\mathalpha}{cmletters}{"76}

\RequirePackage[normalem]{ulem}
\RequirePackage{color}\definecolor{RED}{rgb}{1,0,0}\definecolor{BLUE}{rgb}{0,0,1}\definecolor{GREEN}{rgb}{0,0.5,0}

\title[Disk Formation in Tidal Disruptions]{Tidal Disruption Disks Formed and Fed by Stream-Stream and Stream-Disk Interactions in Global GRHD Simulations}

\author[Zachary L. Andalman et al.]{
Zachary L. Andalman,$^{1,2,3}$\thanks{E-mail: zack.andalman@yale.edu}
Matthew T.P. Liska,$^{4,5}$
Alexander Tchekhovskoy,$^{3}$
\newauthor 
Eric R.~Coughlin$^{6,7}$
and Nicholas Stone$^{8,9,10}$
\\
$^{1}$Yale University, New Haven, CT 06520, USA\\
$^{2}$Evanston Township High School, 1600 Dodge Avenue, Evanston, IL 60201\\
$^{3}$Center for Interdisciplinary Exploration \& Research in Astrophysics (CIERA), Physics \& Astronomy, Northwestern University, Evanston, IL 60202, USA\\
$^{4}$Institute for Theory and Computation, Harvard University, 60 Garden Street, Cambridge, MA 02138, USA; John Harvard Distinguished Science and ITC Fellow\\
$^{5}$Anton Pannekoek Institute for Astronomy, University of Amsterdam, Science Park 904, 1098 XH Amsterdam, The Netherlands\\
$^{6}$Department of Physics, Syracuse University, Syracuse, NY 13244, USA \\
$^{7}$Department of Astrophysical Sciences, Princeton University, Princeton, NJ 08544 \\
$^{8}$Racah Institute of Physics, The Hebrew University, Jerusalem, 91904, Israel\\
$^{9}$Department of Astronomy, University of Maryland, Stadium Drive, College Park, MD, 20742, USA\\
$^{10}$Columbia Astrophysics Laboratory, Columbia University, New York, NY 10027, USA
}

\date{Accepted XXX. Received YYY; in original form ZZZ}
\pubyear{2020}

\begin{document}
\label{firstpage}
\pagerange{\pageref{firstpage}--\pageref{lastpage}}
\maketitle

\begin{abstract}
When a star passes close to a supermassive black hole (BH), the BH's tidal forces rip it apart into a thin stream, leading to a tidal disruption event (TDE). In this work, we study the post-disruption phase of TDEs in general relativistic hydrodynamics (GRHD) using our GPU-accelerated code H-AMR. We carry out the first grid-based simulation of a deep-penetration TDE ($\beta=7$) with realistic system parameters: a black-hole-to-star mass ratio of $10^6$, a parabolic stellar trajectory, and a nonzero BH spin. We also carry out a simulation of a tilted TDE whose stellar orbit is inclined relative to the BH midplane. We show that for our aligned TDE, an accretion disk forms due to the dissipation of orbital energy with $\sim20$ percent of the infalling material reaching the BH. The dissipation is initially dominated by violent self-intersections and later by stream-disk interactions near the pericenter. The self-intersections completely disrupt the incoming stream, resulting in five distinct self-intersection events separated by approximately 12 hours and a flaring in the accretion rate. We also find that the disk is eccentric with mean eccentricity $e \approx 0.88$. For our tilted TDE, we find only partial self-intersections due to nodal precession near pericenter. Although these partial intersections eject gas out of the orbital plane, an accretion disk still forms with a similar accreted fraction of the material to the aligned case. These results have important implications for disk formation in realistic tidal disruptions. For instance, the periodicity in accretion rate induced by the complete stream disruption may explain the flaring events from Swift J1644+57.
\end{abstract}


\section{Introduction}
\label{sec:introduction}

In recent decades, several very bright flares in galactic nuclei have been observed and interpreted as tidal disruption events (TDEs), which occur when a star is scattered onto a nearly parabolic orbit around a supermassive black hole (BH) with a pericenter inside the tidal radius of the BH \citep{Hills1975, Frank1976, Rees1988}. While these flares are typically discovered from quasi-thermal emission in the soft X-ray \citep{Bade1996, Komossa1999, Saxton2012}, UV \citep{Gezari2006, Gezari2008}, or optical \citep{vanVelzen2011, Gezari2012, Arcavi2014, Holoien2014} bands, they have been observed to emit radiation across the electromagnetic spectrum, from radio synchrotron \citep{Zauderer2011, Alexander2017} to nonthermal hard X-rays and soft gamma rays \citep{Bloom2011, Cenk2012, Brown2015}.

Our current theoretical understanding of the tidal disruption process -- the star's first, terminal pericenter passage -- is largely converged \citep{Lacy1982, Carter1983, Guillochon2013, Mainetti2017}, at least for polytropic stars in Newtonian gravity. More recent simulations have explored how the immediate outcome of disruption depends on stellar spin \citep{Golightly+19a, Kagaya+19}, realistic models of the star's internal structure \citep{Golightly+19b, Ryu+20a}, and general relativistic gravity \citep{gafton15, Tejeda2017, gafton19, Ryu+20b}. However, we do not yet have a first-principles understanding of how, or if, the stellar debris streams are able to form a nearly axisymmetric, or quasi-circular, accretion disk (for a recent review of this problem see \citealt{Bonnerot2020}). Because the bound stellar debris has typical eccentricities $0.99 \lesssim e \lesssim 0.999$ \citep{Stone2013}, an enormous excess of orbital energy must be dissipated for circularization to occur.  

Early work conjectured that most of this energy dissipation arises from relativistic apsidal precession \citep{Rees1988}: as the most tightly bound debris passes through pericenter, its apsidal angle measured in radians precesses by an order-unity amount, causing a large-angle collision with less tightly bound matter that has yet to return to pericenter. The shocks that thermalize bulk kinetic energy at the point of self-intersection offer a plausible mechanism for circularizing returning stellar debris \citep{Hayasaki2013, Bonnerot2016}. However, self-intersection shocks may be less efficient at circularizing the debris for inclined orbits around spinning Kerr BHs. In this regime, nodal precession, due to Lense-Thirring frame dragging may delay the onset of self-intersection by many orbits \citep{Cannizzo1990, Kochanek1994, Guillochon2015, Hayasaki2016}. Additionally, energy dissipation due to self-intersection shocks may be greatly limited for less relativistic orbital pericenters, with small-angle collisions occurring at self-intersection radii near the apocenter of the most tightly bound debris \citep{Dai2015, Shiokawa2015}.

An alternative dissipation site is at the stream pericenter itself, where the recompression of the returning debris generates ``pancake'' or ``nozzle'' shocks \citep{Kochanek1994}. Recently, \citet{Bonnerot2021} performed an in-depth study of the nozzle shock using a two-dimensional simulation of a vertical slice of the stream and found that dissipation at the nozzle shock raises the entropy of the gas by two orders of magnitude. Newtonian hydrodynamic simulations by \citet{Ramirez-Ruiz2009} have shown that this pericenter shock could feasibly circularize the tidal debris. However, these simulations were performed for a BH-to-star mass ratio of $Q=10^3$ and analytic estimates of \citet{Guillochon2014} suggest that pericenter recompression shocks become less efficient at more realistic mass ratios (e.g. $Q = 10^6$). It is also possible that in many TDEs, efficient dissipation is lacking altogether, and the formation of an accretion disk is an inefficient process unfolding over many fallback times \citep{Piran2015}. We discuss further in Section \ref{sec:energy}.

TDE debris circularization and disk formation is a complex physical problem involving a large dynamic range, general relativistic orbital dynamics, the need for accurate treatment of hydrodynamic shocks, and possibly even magnetohydrodynamic (MHD) effects \citep{Svirski2017}. The many pieces of multiscale and nonlinear physics involved in TDE disk formation mean that, for numerical reasons, almost all past simulations of this process employed major simplifying assumptions that cast doubt on the generality of their conclusions. 

\citet{Ayal2000} initiated the numerical study of TDE circularization using a post-Newtonian (PN) potential to simulate the lowest-order level of apsidal precession in a finite-mass, smoothed particle hydrodynamics (SPH) framework, albeit with low ($N \sim 10^3$) particle number. More recently, global circularization simulations achieved much higher resolution by reducing the dynamic range of the problem in one of two ways. The first is to consider an unrealistically low mass ratio, typically $Q \sim 10^3$. In simulations of this type, general relativity is sometimes ignored completely \citep{Guillochon2014}, but when it is included, it has a minimal effect on the circularization process because the tidal radius around an intermediate-mass BH is not very relativistic \citep{Ramirez-Ruiz2009, Shiokawa2015}. 

The second option is to consider a realistic mass ratio ($Q \sim 10^6$) but an unrealistic pre-disruption stellar orbit. Tidally disrupted stars typically approach supermassive BHs on nearly parabolic orbits \citep{Magorrian1999}, with initial eccentricities $1-e_0 \sim 10^{-5}$. For computational convenience, one may choose an unrealistic stellar eccentricity, $e_0 \lesssim 0.95$, to reduce the debris stream apocenters. This approach was adopted by \citet{Hayasaki2013}, who mimicked apsidal precession effects with a pseudo-Newtonian potential in a highly relativistic $\beta=5$ TDE. They found rapid circularization due to orbital energy dissipation at stream self-intersections. Later simulations found that in less relativistic $\beta = 1$ TDEs, self-intersections are less efficient at energy dissipation compared to this initial work \citep{Bonnerot2015}.

The results of \citet{Hayasaki2013} were later confirmed and extended to different gas equations of state \citep{Bonnerot2016, Hayasaki2016}, as well as higher (but still sub-parabolic) eccentricities \citep{Bonnerot2016, Sadowski2016}. The low-$e_0$ limit of tidal disruption has also been used with PN potentials to include Lense-Thirring frame dragging, which was seen to substantially delay circularization provided debris streams remain thin \citep{Hayasaki2016}. More recently, \citet{Bonnerot2019} have performed a TDE disk formation simulation with realistic astrophysical parameters using a different approximation: neglecting the returning debris streams entirely, and injecting mass, momentum, and energy (in the form of SPH particles) from the test-particle self-intersection point. The validity of this approach depends on the accuracy of the local injection scheme, and its independence from global gas evolution around the BH. We discuss this approach further and compare and contrast it to our results in Section \ref{sec:BonLu}.

In this paper, we use novel numerical techniques to capture the disk formation process in general relativistic hydrodynamics without sacrificing astrophysical realism in our choice of system parameters (e.g., $Q$, $e_0$). We use two-level adaptive mesh refinement (AMR) to resolve the relevant physics within our grid-based code. In \S \ref{sec:numerics}, we outline our numerical scheme. In \S \ref{sec:results}, we describe the general outcomes of our simulation, including the spatial properties of the nascent accretion flow. In \S \ref{sec:discussion}, we more carefully analyze the specific physical mechanisms controlling the accretion and circularization process and provide a detailed comparison to the ZEro-BeRnoulli Accretion (ZEBRA) model of \citet{Coughlin2014}. We conclude in \S \ref{sec:conclusions}.  

\section{Numerical Method and Setup}
\label{sec:numerics}

We simulate the initial tidal disruption using the SPH code {\sc phantom} \citep{Price2007a} and we simulate the post-disruption evolution using our new GRMHD code H-AMR \citep{Liska2019d}, an approach analogous to those of \citet{Rosswog2009} and \citet{Sadowski2016}. With this hybrid method, we can account for the large range of spatial and temporal scales involved in the disruption process and debris stream formation while accurately capturing the essential shocks and general relativistic effects in the post-disruption evolution. 

\subsection{Initial disruption in {\sc phantom}}
\label{sec:phantom}

The stellar disruption is initially followed by the smoothed-particle hydrodynamics code {\sc phantom} \citep{Price2018}. The setup is identical to that described in \citet{Coughlin2015}: a star of mass $1M_{\odot}$ is modeled as a $\gamma = 5/3$ polytrope, with the adiabatic index equal to the polytropic index, by placing $\sim 10^{7}$ particles on a close-packed sphere. The sphere is stretched to achieve roughly the correct polytropic density profile. The star is subsequently relaxed in isolation (i.e. without the external gravitational potential of the BH) for ten sound crossing times to smooth out numerical perturbations in the density profile. Self-gravity is included through the implementation of a tree algorithm alongside an opening angle criterion to calculate short-range forces \citep{Gafton2011}. We also include the effects of shock heating in modifying the internal energy of the gas. 

The relaxed polytrope is placed at a distance of $5\,r_{\rm t}$ from the supermassive BH of mass $10^{6}M_{\odot}$ such that the center of mass is on a parabolic orbit. To maintain hydrostatic balance initially, every particle comprising the star is given the velocity of the center of mass. In its current version, {\sc phantom} is a Newtonian code, and therefore has no direct means of implementing general relativistic effects. Instead, we mimic some of these effects with a pseudo-Newtonian ``Einstein'' potential used by \citet{Nelson2000, Nealon2015, Xiang2016}. The potential is given by
\begin{equation}
    \Phi_{\rm E} = -\frac{GM_{\rm BH}}{r}\left(1+\frac{3R_{\rm g}}{r}\right), 
    \label{eq:phiE}
\end{equation}
\noindent
where $R_{\rm g} = GM_{\rm BH}/c^2$ is the gravitational radius and $M_{\rm BH}$ is the mass of the BH. This potential accurately reproduces the general relativistic apsidal precession angle at large radii relative to the gravitational radius, with deviations from the true precession angle becoming more pronounced as the radius $r$ becomes comparable to $R_{\rm g}$. However, for the large mass ratio considered here, the tidal approximation is upheld to a high degree of accuracy, meaning that the dominant effect of general relativity on the initial stellar encounter will be to rotate the entire star through the same precession angle\footnote{The tidal tensor implied by the Einstein potential produces a moderately different tidal shear than the exact general relativistic value. However, the general relativistic corrections to the Newtonian tidal shear (either exact or from Equation \ref{eq:phiE}) are only $\mathcal{O}(5\%)$ at the tidal radius where the star is disrupted, and ballistic motion sets in.}. Therefore, our usage of this potential, as opposed to a fully general relativistic treatment, is sufficient for the purpose of creating a realistic distribution of post-disruption debris. See \citet{Tejeda2013} for a detailed evaluation of a similar potential used by \citet{Nowak1991}.

The initial, parabolic orbit of the star is established using the above potential (Equation \ref{eq:phiE}) to calculate the angular momentum necessary to achieve a pericenter distance of $r_{\rm p} = 7 R_{\rm g}$. {\sc phantom} uses an artificial viscosity prescription to mediate any strong shocks that may be present during the large compression suffered by the star and employs the standard switch proposed by \citet{Cullen2010} (i.e. the artificial viscosity parameter is small when the star is far from pericenter and approaches values near unity as the star is compressed at pericenter). A nonlinear term is also included to account for extremely strong shocks and prevent interparticle penetration \citep{Price2010}. The large number of particles ($\sim 10^{7}$) was used to avoid the possibility of spurious numerical heating at pericenter caused by under-resolving the compression, predicted to be of the order $H_{\rm min}/R_* \sim \beta^{-3} \sim 0.003$ (\citealt{Carter1983}), though the compression could be smaller if shock heating halts the otherwise-adiabatic collapse (\citealt{Bicknell1983}), or due to three-dimensional effects \citep{Guillochon2009}. Here $\beta = r_{\rm t}/r_{\rm p}$ is the penetration factor, $r_{\rm t}$ is the tidal radius and $r_{\rm p}$ is the pericenter radius.

\begin{figure}
	\includegraphics[width=1.0\columnwidth]{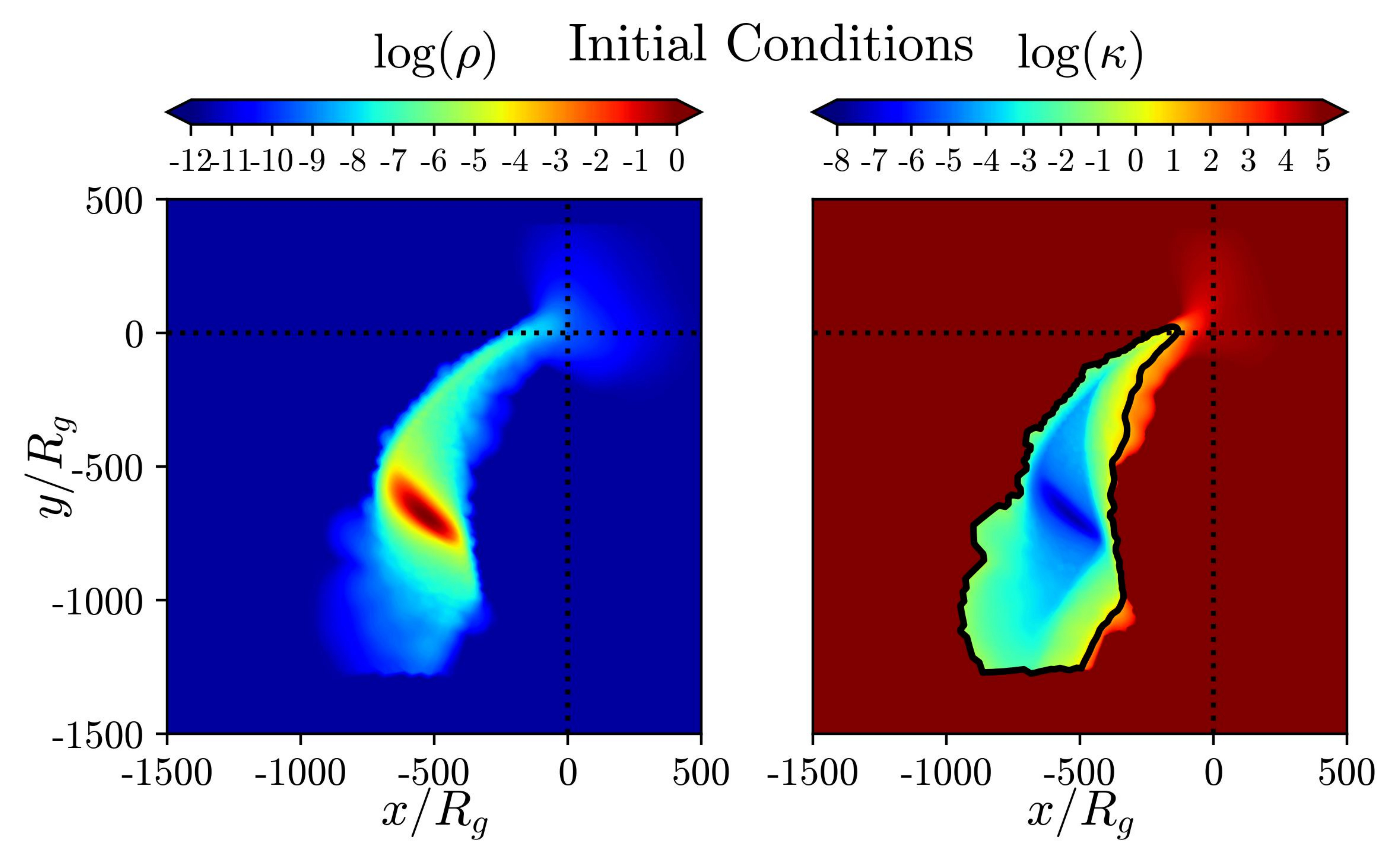}
    \caption{Color maps of the log of rest mass density and $\log\kappa$ (proportional to entropy, see Equation \ref{eq:kappa}), in the equatorial plane at the initial conditions of the post-disruption phase of the simulation in H-AMR (1.16 days). The black contour on the right panel outlines the area excluded by the entropy condition ($\kappa$ < 10) which we use throughout our analysis to distinguish the material in the debris stream from the material in the accretion disk). The BH is located at the origin. The dotted lines indicate the $x$- and $y$-axes.}
    \label{fig:init}
\end{figure}

\begin{figure}
	\includegraphics[width=1.0\columnwidth]{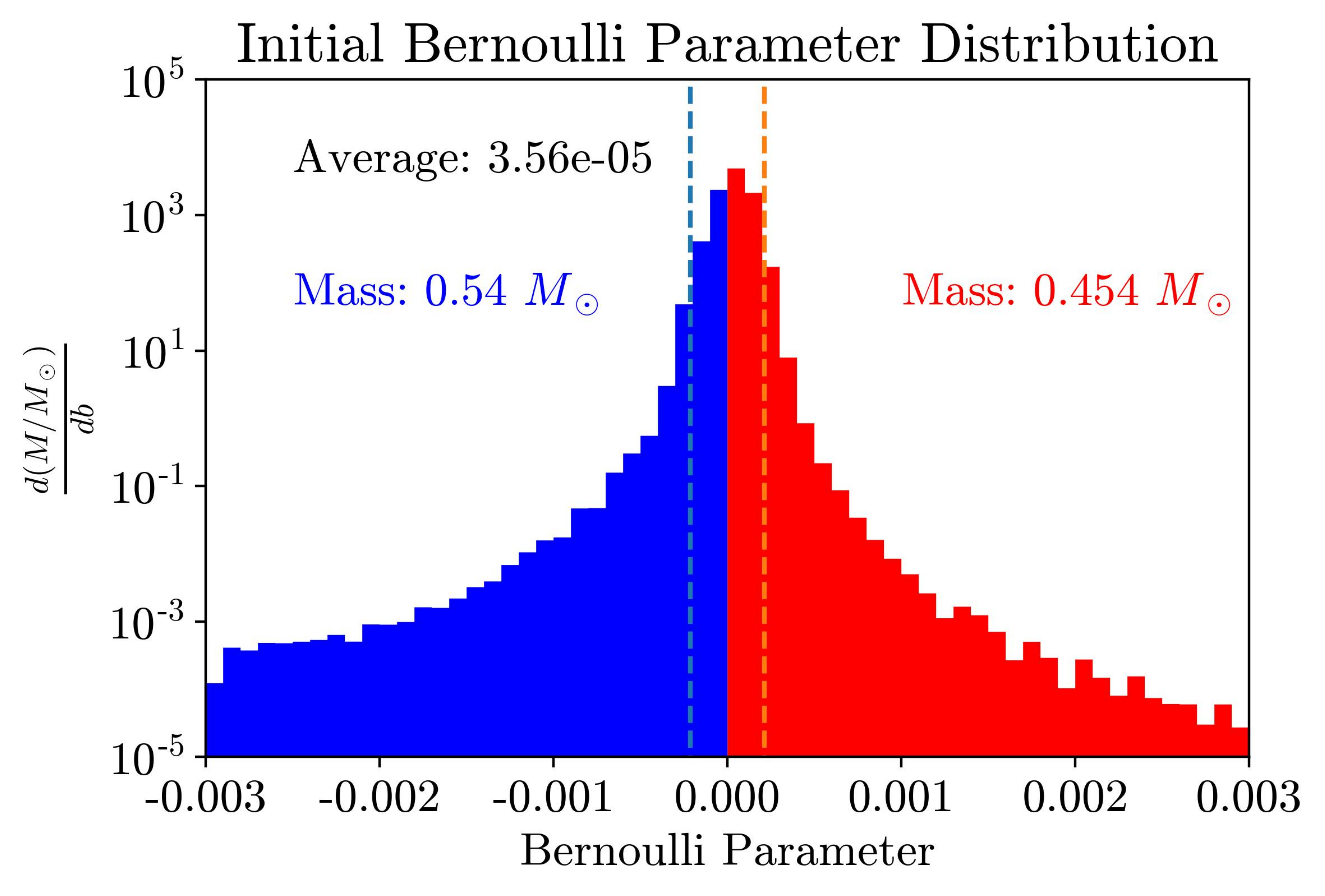}
    \caption{A histogram of the Bernoulli parameter distribution at the initial conditions of the post-disruption phase of the simulation in H-AMR (1.16 days). Each bin is weighted by solar masses and bin width. Unbound material and total unbound mass are shown in red; bound material and total bound mass are shown in blue. The mass-weighted average Bernoulli parameter is also shown. On average, the material in the initial conditions of the post-disruption phase is marginally bound ($b=0$). The vertical lines represent the range of the Bernoulli parameter estimated from the frozen-in approximation (Equation \ref{eq:frozen}), and contain 98.1 percent of the debris mass. The floor material is ignored using a density condition ($\rho > 10^{-11}$).}
    \label{fig:Bernoulli}
\end{figure}

Figure \ref{fig:init} shows the density distribution of the disrupted stellar debris at 1.16 days after the disruption. At this time, we end the evolution of the TDE in {\sc phantom} and use the distribution of debris as the initial conditions for our post-disruption simulation in H-AMR. See Section \ref{sec:mapping} for a detailed description of how we map data from SPH to grid-based GRHD.

Figure \ref{fig:Bernoulli} depicts the Bernoulli parameter of the tidally-disrupted debris at this same time. The star approaches the BH on a parabolic orbit. When the BH tidally disrupts the star, the bound stellar debris falls back to the BH while the unbound debris continues on an outward trajectory. The star's mass is split almost evenly between bound and unbound matter \citep{Lacy1982, EvansKochanek1989}. We use the relativistic Bernoulli parameter to distinguish between bound and unbound material.

\begin{equation}
    b=-\frac{u_t(\rho c^2 + u_g \gamma)}{\rho c^2}-1
    \label{eq:Bernoulli Parameter},
\end{equation}

\noindent
where $b$ is the Bernoulli parameter, $u$ is the 4-velocity, and $\rho$ and $u_g$ are the mass and internal energy densities in the fluid frame. At late times the pressure gradient within the disrupted debris becomes small and the Bernoulli parameter is approximately a conserved Lagrangian quantity. Even in a time-dependent system such as the one analyzed here, $b>0$ corresponds to unbound material while $b<0$ corresponds to bound material. 

Figure \ref{fig:Bernoulli} shows that in the initial conditions, the majority of the material is marginally bound with Bernoulli parameter inside the range predicted by the frozen-in approximation of \citet{Stone2013}. This approximation assumes that (i) the tidal forces outside of the tidal radius are negligible, so the star enters the tidal radius as an unperturbed sphere, and (ii) once the star crosses the tidal sphere, its fluid elements move ballistically, with a spread in orbital properties given by the potential gradient across the star. In reality, internal forces (e.g. self-gravity and hydrodynamics) are not totally negligible inside the tidal sphere, but previous simulations of deeply penetrating disruptions show that the frozen-in approximation reproduces the actual energy spread of the debris to within  $\approx 20$ percent for $\gamma=5/3$ polytropes \citep{steinberg19}. According to this impulsive disruption approximation, the spread of specific orbital energy $\Delta b$ in Newtonian gravity is given by

\begin{equation}
    \Delta b = k \frac{GM_{\textup{BH}}R_*}{r_{\rm t}^2}
    \label{eq:frozen}
\end{equation}

\noindent
where $k$ is a constant of order unity related to stellar structure and rotation prior to disruption. If we let $k=1$, we find that $\Delta b = 2.12 \times 10^{-4}$. Only 1.94 percent of the mass in the initial conditions is outside the range predicted by the frozen-in approximation, verifying that the initial orbital energy distribution for the post-disruption phase is largely consistent with standard estimates.

A small fraction of the material has Bernoulli parameters well outside the range predicted by the frozen-in approximation. However, even though the most tightly bound debris (with specific energy $|\varepsilon| > \Delta b$) constitutes a small fraction of the total mass, it is the first matter to fall back, and therefore dominates the early stages of the circularization process studied here. Due to runtime limitations, these early stages are the primary focus of this paper. While these tails could be a byproduct of intense shock heating as the star is highly compressed near pericenter, we caution that they may also arise from numerical inaccuracies associated with the same highly compressed, and therefore difficult to resolve, configuration of gas.  

Such broad-energy tails have been seen in high-$\beta$ TDEs simulated with a range of codes and numerical algorithms. While a return time of $1.16$ days for the most tightly bound debris might appear extreme, it is qualitatively consistent with these past simulations. For example, \citet{Guillochon2013} find that the most tightly bound debris in Newtonian, grid-based, $\beta=4$ simulations of $n=3$ polytrope disruptions can return to pericenter after $\approx 3$ days and that the time of first pericenter return decreases with increasing $\beta$. \citet{steinberg19} performed moving-mesh simulations of stellar disruptions and found that the extent of the high energy tail is also a function of $\beta$. For Newtonian disruptions of $n=3/2$ polytropes, going from $\beta=5$ to $\beta=7$ moves the time of first mass return from $\approx 3$ days to $\approx 1$ day (Steinberg, private communication). \citet{gafton19} used Newtonian and relativistic SPH simulations, with a code distinct from {\sc phantom}, to disrupt a $\gamma = 5/3$ polytrope over a range of $\beta$ and found that for large $\beta$ the return time of the most bound debris was significantly earlier than the frozen-in prediction, with initial return times on the order of days. 

These high-energy, low-mass debris tails have not been studied in detail, but their ubiquity across SPH, conventional grid-based, and moving mesh codes leads us to believe that they are likely physical. If, however, the high-energy tail of debris were primarily the result of numerical artifacts, then it would bias the earliest stages of mass return to (i) artificially early times and (ii) artificially low fallback rates relative to the time of first mass return.

However, our results depend solely on the relative values of mass fluxes rather than the absolute values, with the exception of the internal energy and density floors (Section \ref{sec:params}). Therefore, even if the mass of the high-energy tail of debris in our initial conditions is an overestimate, our results can be straightforwardly rescaled to astrophysically realistic time and mass flux scales (i.e. our simulation would have started at a later time with similar values of relative mass flux). The qualitative features of the circularization process are therefore robust and should apply generically to systems with realistic physical parameters and $\beta \simeq 7$.

\subsection{Mapping from {\sc phantom} to H-AMR}
\label{sec:mapping}

Before we begin our simulations in H-AMR, we map SPH data from {\sc phantom} to gridded data compatible with GRHD. First, smooth the particle properties onto a continuous domain. Second, we construct a grid on top of the smoothed particles.

For the first step, we use the SPH visualization tool {\sc splash} and its built-in function ``splash to grid,'' which is described in detail in \citet{Price2007b}. The function smooths the density of each particle over a finite region according to a weighting function, or kernel, that is twice differentiable, maintains compact support, and decreases in magnitude from the location of the particle. This approach mirrors the standard procedure for SPH calculations. We use the default kernel in {\sc splash} (and {\sc phantom}): a cubic spline which vanishes at a distance of $2h$ from a given particle. The smoothing length $h$ is spatially variables and set such that the mass inside the smoothing sphere is constant. We refer the reader to Section 2.4 of \citet{price12} for more details.

For the second step, the fluid variables of a given cell are determined by adding the contribution of every particle with a smoothing region that encompasses the cell itself. See Figure 7 of \citet{Price2007b} for an illustration. This method ensures that cells in high-density regions are sampled by a large number of particles and cells in low-density regions are sampled by relatively few particles. Because of their low density, sparsely sampled cells contribute minimally to the dynamics of the fluid and do not affect the bulk properties of the accretion flow simulated with H-AMR.

The error in the first step is of order $\mathcal{O}(h^2) = \mathcal{O}(\rho^{-2/3})$. Therefore, in regions of extremely low density, the interpolation from SPH may introduce artifacts in the density and velocity fields. In particular, we find a region of enhanced density in the region around the black hole (Figure \ref{fig:init}). The density of this region is several orders of magnitude lower than the density of the most bound debris, so the trajectory of the returning debris is not significantly altered.

\subsection{H-AMR Simulation Parameters}
\label{sec:params}

\begin{table}
	\centering
	\caption{Simulation parameters for models TDET0 and TDET30, including black hole mass $M_{\rm BH}$, stellar mass $M_*$, pericenter radius $R_p$, penetration factor $\beta$, and inclination angle of the stellar orbit $i$.}
	\label{tab:models}
	\begin{tabular}{lccccr}
		\hline
		Model & $M_{\rm BH}$ ($M_\odot$) & $M_*$ ($M_\odot$) & $R_p$ ($R_g$) & $\beta$ & $i$\\
		\hline
		TDET0 & $10^6$ & $1$ & $7$ & $7$ & $0$\\
		TDET30 & $10^6$ & $1$ & $7$ & $7$ & $30$\\
		\hline
	\end{tabular}
\end{table}

As described in Section \ref{sec:introduction}, our simulation parameters adhere to astrophysically realistic values ($Q=10^6$, $e_0 \approx 1$). In {\sc phantom} we insert a star on a parabolic trajectory with a pericenter distance of $7 R_{\rm g}$ and a penetration factor of $\beta=r_{\rm t}/r_{\rm p}=7$. This high penetration encounter guarantees that self-gravity is negligible in the post-disruption evolution of the stream, although the influence of self-gravity on the stream structure may be revived at much later times than those simulated here due to the in-plane focusing of the debris; \citealt{coughlin16, steinberg19}). The circularization of the stellar debris likely occurs on a shorter timescale for more relativistic encounters \citep{Bonnerot2020}, decreasing the simulation duration required to study the circularization process.

H-AMR uses a naturalized unit system where $G=c=R_{\rm g}=1$. The conversion factors from the simulation units to cgs units are given in Table \ref{tab:units}. From now on, we will work in this naturalized unit system with the exception of time, which we will convert back to physical units of days since disruption. We will also restore $G$ and $c$ in equations to help keep track of units.

\begin{table}
	\centering
	\caption{For each quantity, the number of cgs units per simulation unit is tabulated. Note that 0 $R_{\rm g}/c$ corresponds to 1.16 days after the disruption, so the relationship between simulation time and days since disruption is affine linear.}
	\label{tab:units}
	\begin{tabular}{lccr}
		\hline
		Quantity & cgs unit & H-AMR unit & cgs unit / H-AMR unit\\
		\hline
		Mass & g & $R_{\rm g} c^2/G$ & $2 \times 10^{39}$\\
		Distance & cm & $R_{\rm g}$ & $1.477 \times 10^{11}$\\
		Time & s & $R_{\rm g}/c$ & $4.926$\\
		Density & g/cm$^3$ & $c^2/R_{\rm g}^2 G$ & $6.207\times 10^5$\\
		\hline
	\end{tabular}
\end{table}

Although H-AMR does not explicitly include viscosity, it is included implicitly through interactions at the cell level. Within the turbulent flows of our simulation, adjacent fluid elements are unlikely to move exactly parallel to one another and therefore will exchange momenta. Previous work has shown that the effective viscosity in early-time TDE accretion flows can be dominated by the Reynolds stress \citep{Sadowski2016}. If this is correct, we should not be significantly underestimating effective viscosity due to the absence of magnetic fields, though this question needs closer examination in future magnetohydrodynamic simulations.

We present two models, TDET0 and TDET30, corresponding to spin-orbit misalignment angles of zero and 30 degrees, respectively (Table~\ref{tab:models}; see the 3D renderings in the \hyperref[sec:support]{Supporting Information}). Because we begin the simulation in H-AMR $1.16$ days post disruption, the relationship between H-AMR simulation time and time since disruption is affine linear.

\begin{equation}
    t_{\rm HAMR} = (t_{\rm days} - 1.16) \times 24 \times 3600 / 4.926
    \label{eq:time_conversion}
\end{equation}

In both models, we use a dimensionless BH spin of $a=0.9375$ for the post-disruption evolution. Because the morphology and the fallback rate of our stream are not strongly dependent on spin \citep{Tejeda2017}, we rotate the initial the initial data in H-AMR about the $y$-axis by 30 degrees for model TDET30 rather than repeating the simulation in {\sc phantom}. We take this approach, rather than tilting the metric, to avoid the computational strain associated with a non-axisymmetric metric. 

We run models TDET0 and TDET30 until 6.87 days ($t=10^5 R_{\rm g}/c$ in H-AMR) and 5.01 days ($t=6.7 \times 10^4 R_{\rm g}/c$ in H-AMR) after the disruption respectively. We evolve the models in the Kerr geometry using Kerr-Schild coordinates to avoid the coordinate singularity in the Boyer-Lindquist coordinates. 

In this work, H-AMR uses 2-level 3D adaptive mesh refinement (AMR) with a refinement criterion based on a threshold density. The total effective resolution is $2880\times860\times1200$ in $r\times\theta\times\phi$. The cells are logarithmically spaced in the radial direction. The cell dimensions in $R_g$ are given approximately as a function of position around the black hole:
\begin{equation}
    \Delta r \times \Delta \theta \times \Delta \phi \approx \frac{\ln(10^5)}{2880} r \times \frac{\pi}{860} r \times \frac{2\pi}{1200} r\sin \theta.
\end{equation}
Our grid resolves the vertical extent of the stream (twice the stream scale height) with $\sim$12 cells near 500 $R_{\rm g}$ and $\sim$28 cells near pericenter, where the stream scale height is computed via the same method as the disk scale height discussed in Section \ref{sec:props}. However, near pericenter, this calculation is artificially inflated by the stream structure's complex $\phi$-dependence. This resolution is sufficient to model the gas dynamics of the stream. However, we caution that we may underresolve pressure gradients within the stream.

In addition to AMR, H-AMR uses local adaptive time stepping by setting the time step in each cell to the smallest light crossing time in that cell. As a result, the timesteps decrease by a factor of $\Delta \phi / \Delta \theta$ near the pole due to the time step limitation in $\varphi$. Therefore, we use the cylindrification method described by \citet{Tchekhovskoy2011} to reduce the azimuthal extent of cells near the pole.

As we ran the simulation, we noticed that the stream disintegrates into the accretion disk after wrapping around the BH, a behavior which we discuss more in Section \ref{sec:props}. To verify that this stream disintegration was not a numerical artifact, we adjusted the refinement criterion for first-order refinement between 2.28 days ($4\times 10^4 R_{\rm g}/c$) and 4.56 days ($8\times 10^4R_{\rm g}/c$). Near the BH, we decreased the cutoff density for first-order refinement to achieve the full effective resolution in a greater fraction of cells. Due to memory restrictions, we also had to increase the cutoff density at large radii, causing some parts of the outer stream to become unrefined. We found that the stream disintegration and other simulation properties were consistent across adjustments to the refinement criterion, suggesting that the physics converged for our resolution in H-AMR.

In the post-disruption phase, we set floors for internal energy density and mass density at $2.27\times10^{-12}$ ($3.75\times 10^6$ ergs/cm$^3$ and $4.167\times 10^{-15}$ g/cm$^3$ respectively). We assume an adiabatic index $\gamma=5/3$ corresponding to the gas pressure dominated regime present in the star before it undergoes shocks. Although any accretion disk resulting from the TDE is expected to be radiation pressure dominated with an adiabatic index closer to $\gamma=4/3$, H-AMR currently does not allow for a variable adiabatic index so we chose to stay consistent with the initial simulation in {\sc phantom}. Many previous works have also used an adiabatic index of $\gamma=5/3$ \citep{Guillochon2015, Sadowski2016, steinberg19}. In future simulations, we will implement a variable adiabatic index to more accurately model the thermodynamics.

\section{Results}
\label{sec:results}

\subsection{Aligned Disk Formation and Evolution}
\label{sec:props}

\begin{figure*}
	\includegraphics[width=1.0\linewidth]{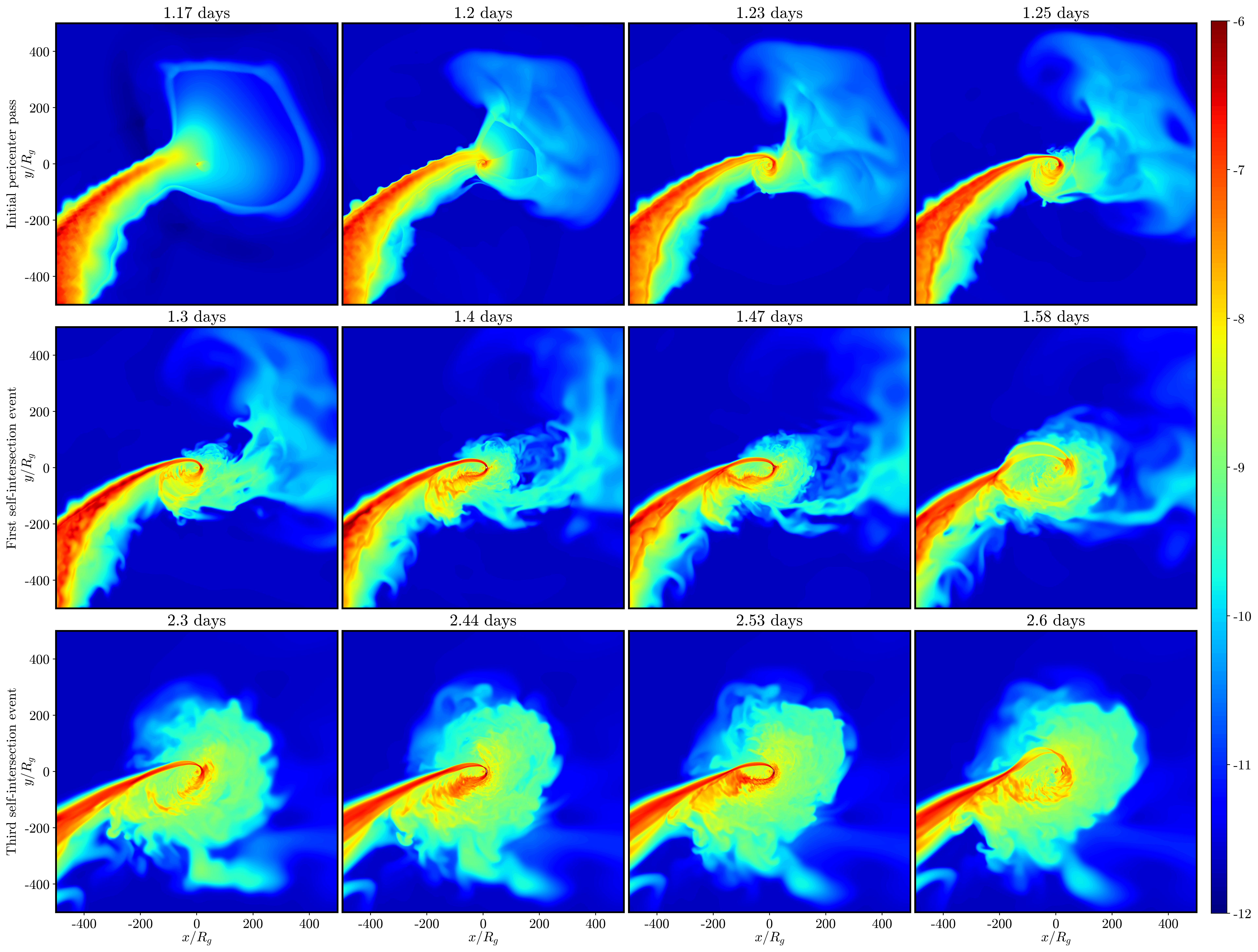}
    \caption{Contour plots of the log of rest mass density in the equatorial plane in the TDET0 simulation during the debris' initial pericenter passage (first row) and the first (second row) and third (third row) self-intersection events. In the initial pericenter passage, the stream falls back through near vacuum and matter from the stream begins to accumulate near the BH. In the self-intersection events, the stream undergoes apsidal precession and self-intersects close to the analytical self-intersection radius at $142 R_{\rm g}$. As a result, the inner parts of the stream are completely disrupted. These violent events are a key dissipation mechanism in the early stages of the TDE evolution. Although powerful, we count only 5 such events. At late times in our simulation, dissipation occurs primarily through interactions with the newly formed disc (Figure \ref{fig:dissolve}). For a more complete picture of the disk evolution, see the movies and 3D renderings linked in Section \ref{sec:support}.}
    \label{fig:intersect}
\end{figure*}

General relativistic apsidal precession near the BH causes the argument of pericenter to advance by approximately $31^\circ$, setting the incoming and outgoing streams on trajectories that intersect at $R_{\rm SI} = 142 R_g$. The precession angle and self-intersection radius are computed with the formalism in Appendix \ref{sec:Rsi}.

\begin{figure}
    \centering
    \includegraphics[width=1.0\columnwidth]{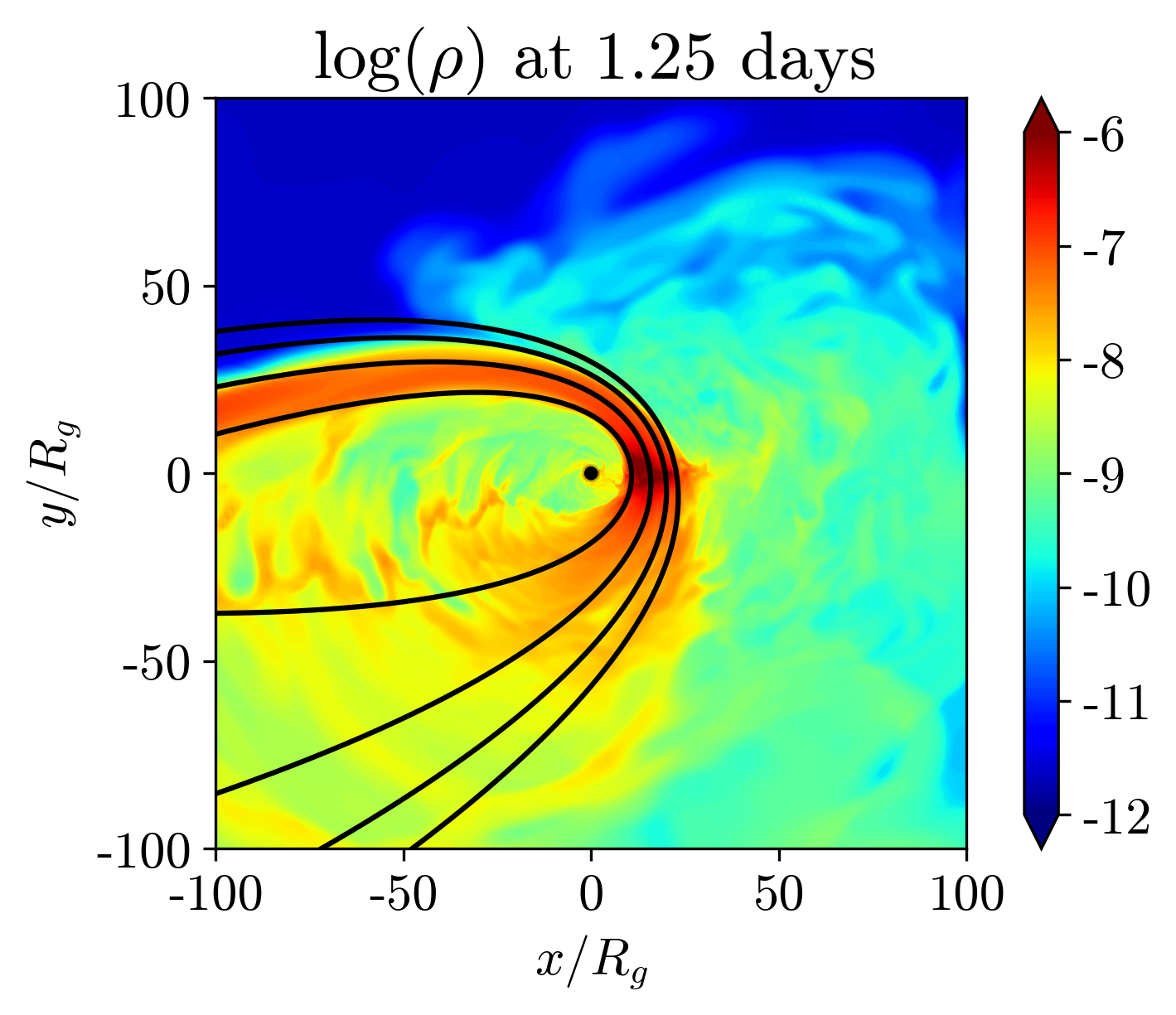}
    \caption{A contour plot of the log of rest mass density in the equatorial plane at 1.25 days ($1725 R_{\rm g}/c$; same as the top-right panel of Figure \ref{fig:intersect}) with equatorial geodesics of various pericenter radii. The geodesics diverge after pericenter, demonstrating the strong effect of differential precession in expanding the stream in the deeply penetrating ($\beta = 7$) and high-spin ($a = 0.9375$) encounter we consider.}
    \label{fig:diff_prec}
\end{figure}

Shortly after the first pericenter passage, depicted in the first row of Figure \ref{fig:intersect}, the debris stream expands significantly and its density drops. The primary reason for this expansion is the different degrees of apsidal precession experienced by the inner- and outer-most edges of the stream. For instance, at $t = 1.25\ {\rm days}$ (top-right panel of Figure \ref{fig:intersect}), the inner and outer edges of the stream at pericenter lie at $r=11 R_g$ and $r=21 R_g$ respectively, resulting in a difference of roughly $15^\circ$ in their precession angle (Figure \ref{fig:diff_prec}). \citet{Bonnerot2021} discuss this effect as being potentially responsible for a small amount of spreading of the debris following its first return to the initial point of disruption. Here, because the pericenter distance of the star is much more relativistic than the modest-$\beta$ case that they considered, the differential precession is larger than they find and apparent by eye.

Over the course of the simulation, the self-intersection periodically becomes powerful enough to fully intercept the incoming debris stream. During these violent self-intersection events, the collision of the incoming and outgoing streams and the associated shock heating nearly destroys the stream interior to the self-intersection point and ejects material into a wide range of orbits. Rows two and three of Figure \ref{fig:intersect} depict the time evolution of two such intersection/depletion cycles.

We find five violent self-intersection events in our simulation. They occur approximately 12 hours apart (at 1.47 days, 2.01 days, 2.52 days, 2.92 days, and 3.68 days), and each lasts for roughly 2.74 hours ($2,000 R_{\rm g}/c$). The periodicity of the self-intersections is on the scale of the free-fall time from the self-intersection point,
\begin{equation}
    t_{\rm{ff}}=\frac{\pi}{2}\frac{R_{\rm SI}^{3/2}}{\sqrt{2GM_{\textup{BH}}}}\approx 3.64\  \rm{hr}.
    \label{eq:freefall}
\end{equation}
These violent, periodic self-intersections events may produce natural, quasi-periodic variability in the inner disk accretion rate, possibly explaining the flaring events observed in TDEs such as SWJ1644+57 \citep{Burrows2011, Zauderer2011}, AT2018fyk \citep{Wevers+19}, and AT2019ehz \citep{vanVelzen+20}. We discuss this hypothesis further in Section \ref{sec:accretion}. The self-intersection events also create the initial accretion disk. 

\begin{figure*}
	\includegraphics[width=1.0\linewidth]{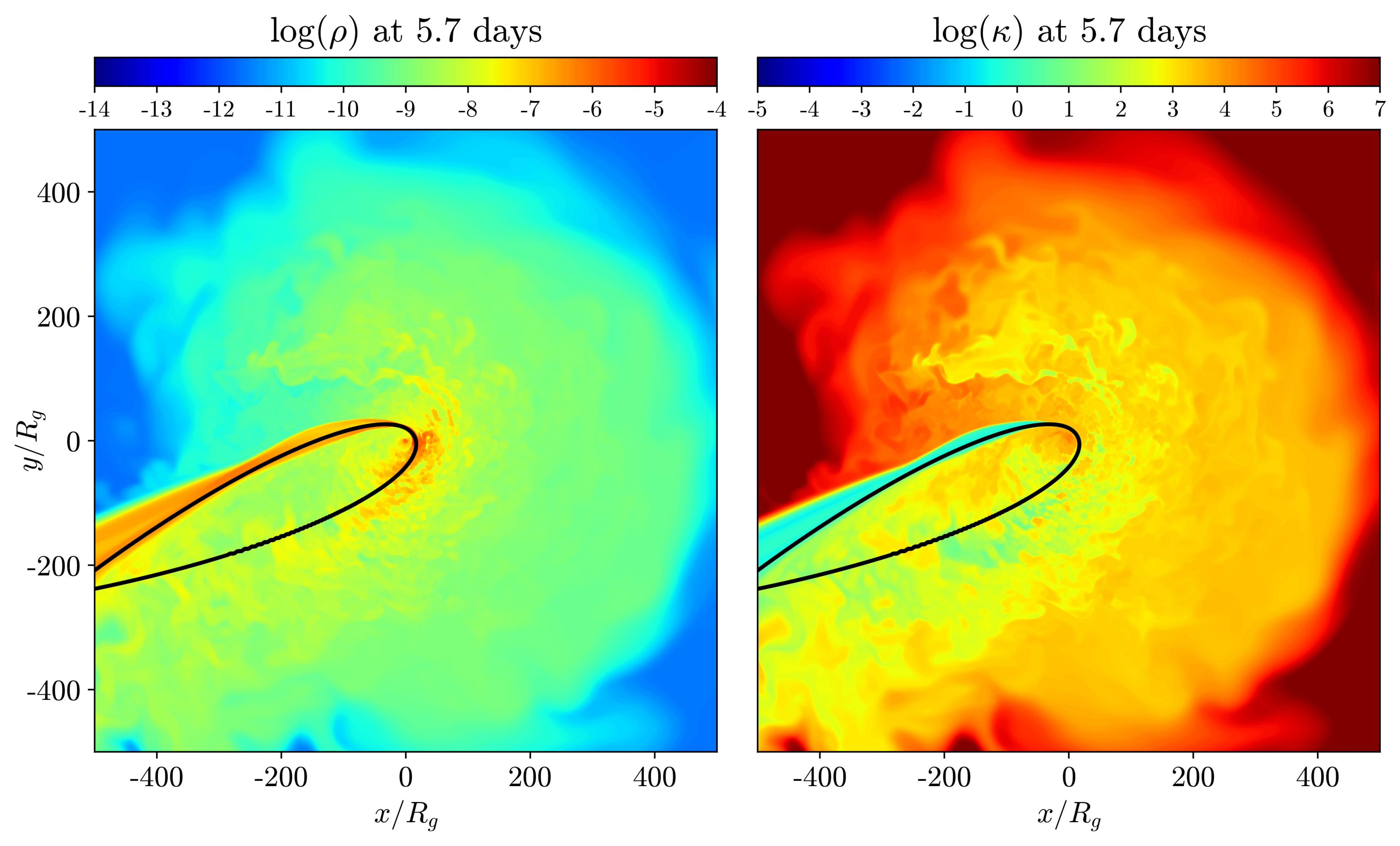}
    \caption{Contour plots of the log of rest mass density density (left panel) and $\kappa$ (right panel) in the equatorial plane at 5.7 days. At late times in the simulation, the debris stream undergoes shocks and instabilities near pericenter, causing it to disintegrate shortly after the pericenter passage. This is different from the stream's early time evolution (Figure \ref{fig:intersect}), possibly because an inner accretion disk has formed with density comparable to the incoming stream. The black line depicts an equatorial geodesic (Appendix \ref{adx:geo}). Note that the self-intersection radius of the geodesic is much greater than the analytical self-intersection radius of $142 R_{\rm g}$ for a pericenter of $7 R_{\rm g}$ because the geodesic has a larger pericenter radius of $\sim 12 R_{\rm g}$.}
    \label{fig:dissolve}
\end{figure*}

However, once the accretion disk becomes sufficiently dense and massive, no additional violent self-intersections occur because the outgoing stream completely disintegrates before the self-intersection point as in Figure \ref{fig:dissolve}. At late times, the spread in angular momenta within the incoming stream grows due to angular momentum exchange with the accretion disk. As a result, the stream becomes thicker, enhancing the effect of differential precession discussed earlier in this section. The result is that the outgoing stream has a larger spread in trajectories and a lower density than the incoming stream. At even later times, the disk is sufficiently dense to absorb nearly all of the momentum from the weakened outgoing stream through shocks and hydrodynamic instabilities at the interface. In particular, the velocity difference between the outgoing stream and the disk leads to a Kelvin-Helmholtz instability (see \citet{Bonnerot2020} Section 2.2.4), which seeds turbulence in the outgoing stream causing it to break apart.

At early times, especially before substantial disk formation, the pericenter radius undergoes fluctuations. From the movies linked in Section \ref{sec:support}, we see that the outward movements of the pericenter coincide with the onset of a violent self-intersection. This suggests that the pericenter movement results from the azimuthal momentum added to the incoming stream by the outgoing stream during a self-intersection. After the self-intersections cease, the pericenter radius becomes more stable and takes on a value around $12 R_g$ (corresponding to a self-intersection radius $R_{\rm SI} = 565 R_g$). This may be because the outgoing stream has a weaker, but more stable, contribution to the azimuthal momentum of the incoming stream at late times.

Instead of forming a standard, geometrically thin disk, the material surrounding the black hole is inflated into a geometrically thick structure that is both gas-pressure and centrifugally supported. We perform a more in-depth analysis of the force balance in the disk and the disk structure in relation to analytical models in Sections~\ref{sec:force} and \ref{sec:disk}.

We use an entropy cutoff to distinguish between the matter in the stream and the matter in the disk. Throughout the remainder of this work, we use the quantity,
\begin{equation}
    \kappa=p\rho^{-\gamma}=\frac{(\gamma-1)u_g}{\rho^\gamma}
    \label{eq:kappa}
\end{equation}
to track specific entropy, which is related to $\kappa$ by
\begin{equation}
    \mathcal{S}=\frac{\textup{ln} \kappa}{\gamma - 1}.
    \label{eq:Entropy}
\end{equation}

The tidal compression of returning debris streams is approximately a reversible process, so entropy is nearly constant until the first shock. For the purposes of analysis, we define the stream as material with $\kappa<10$, a definition we refer to as the entropy condition. Figure \ref{fig:init} depicts an entropy profile of the stream at the initial conditions of the post-disruption phase in the equatorial slice. The black contour outlines the area covered by the entropy condition.

\begin{figure}
	\includegraphics[width=\columnwidth]{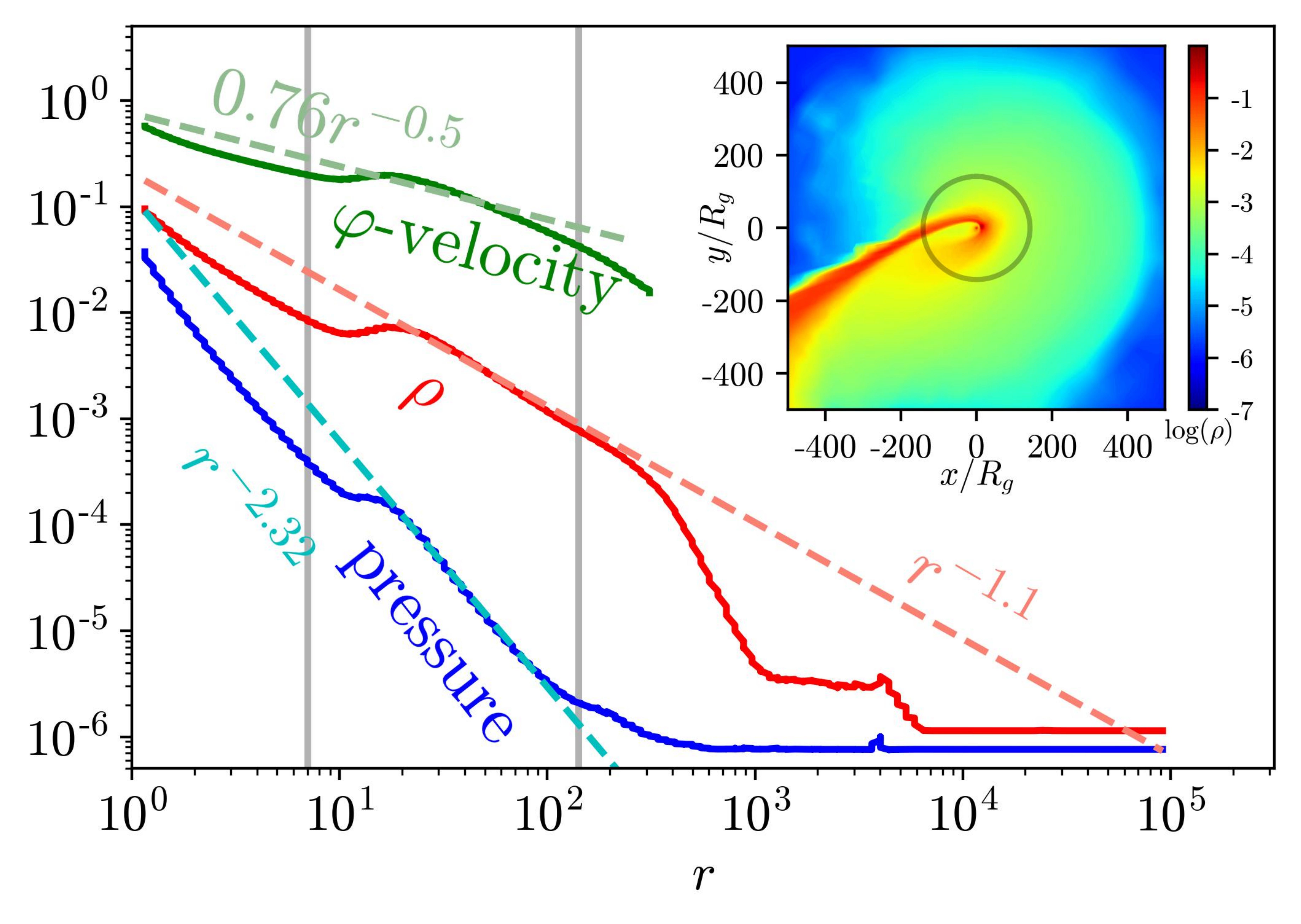}
    \caption{Time averages of radial profiles of mass density, pressure, and $\varphi$-velocity, their power law fits, and an inset plot of time-averaged rest mass density in the equatorial plane. Power law fits (dashed lines) are calculated using a least-squares method (see Table \ref{tab:curve_fit} for more details). Time averages are over the simulation's full duration. Mass density, pressure, and $\varphi$-velocity are averaged over spherical shells using Equation \ref{eq:1davg}. Pressure and $\varphi$-velocity are weighted by mass. The stream is ignored using the entropy condition. Mass, density, and pressure are multiplied by $5 \times 10^5$ so that all three variables are roughly the same order of magnitude for ease of comparison. The vertical lines show the pericenter radius at $7 R_{\rm g}$ and the analytical self-intersection radius at $142 R_{\rm g}$ (Appendix \ref{sec:Rsi}). The analytical self-intersection radius is also shown on the inset plot. All three quantities follow power law fits within the radii of the disk. The sub-unity coefficient on the $\varphi$-velocity indicates that the disk is sub-Keplerian. At radii less than the pericenter radius or greater than 400 $R_{\rm g}$, there is minimal disk material, so the data at these radii does not reflect the large-scale properties of the disk. The origin of the flat density and pressure regions at large radii ($r\gtrsim 1000 R_{\rm g}$) is due to the choice of the floors. The origin of the dips in all quantities at small radii is due to the absence of disk material in the plunging region.}
    \label{fig:radial}
\end{figure}

\begin{figure}
	\includegraphics[width=1.0\columnwidth]{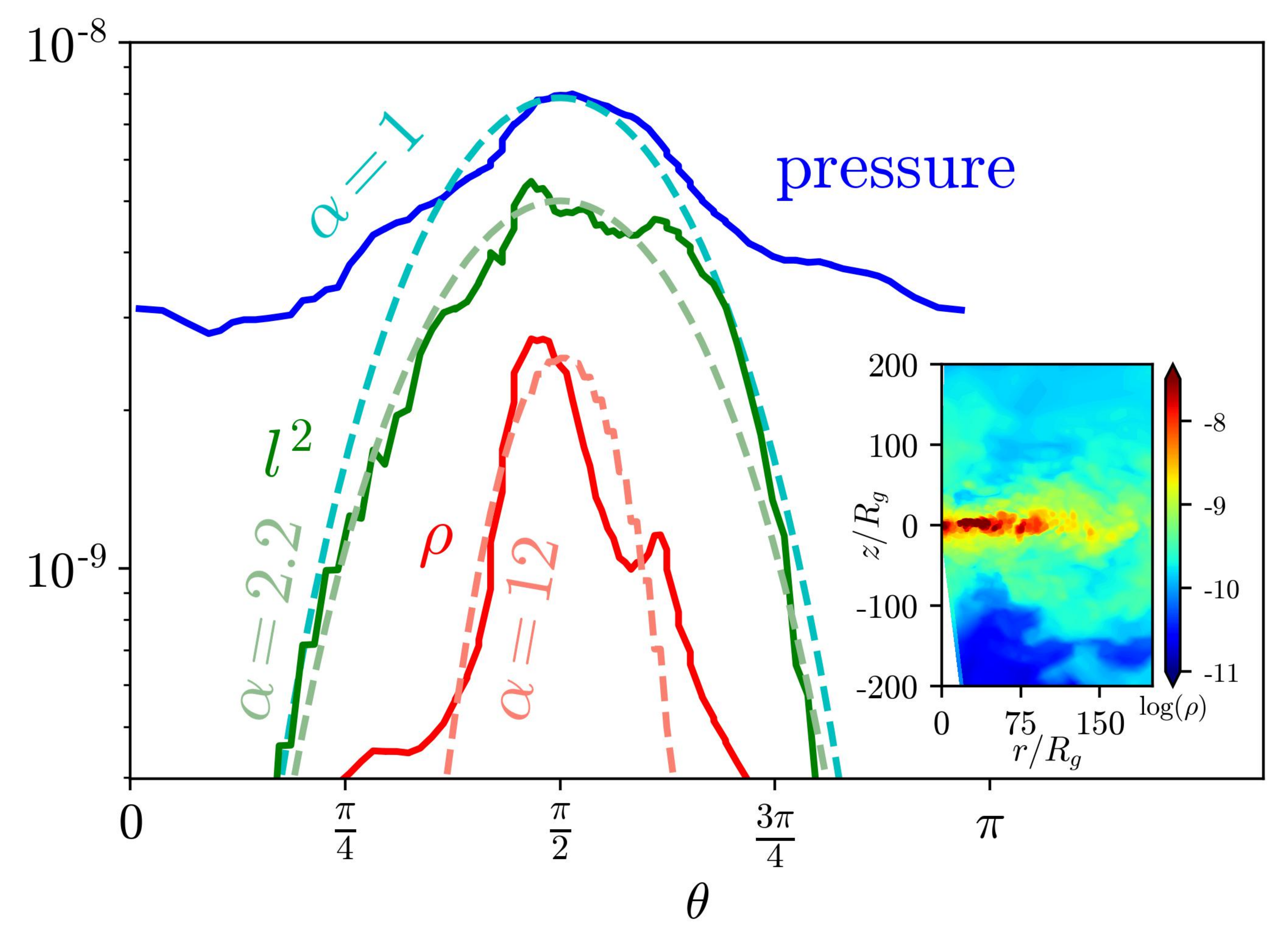}
    \caption{Mass density, pressure, and angular momentum squared in our aligned TDET0 model plotted with respect to $\theta$, their fits to a power law of $\sin^2\theta$, and an inset plot of rest mass density in the $xz$-plane at 5.7 days. Curve fits are estimated by eye and shown in dashed lines. $\alpha$ represents the exponent of a power law of $\sin^2\theta$. Angular momentum is normalized in radius with a factor of $r^{-1/2}$. Density, pressure, and angular momentum are averaged over $\varphi$ and $10 < r < 100$ using Equation \ref{eq:1davg}. Pressure and angular momentum are weighted by mass. The stream is ignored using the entropy condition. Density, pressure, and angular momentum squared are multiplied by $0.158$, $316$, and $10^{-8}$ respectively so that all three quantities are roughly the same order for comparison purposes.}
    \label{fig:polar}
\end{figure}

Figure \ref{fig:radial} shows the radial profiles of density, pressure, and $\varphi$-velocity within the disk. We compute the gas pressure $p$ using the adiabatic equation of state, 
\begin{equation}
    p=u_g(\gamma-1).
    \label{eq:pressure}
\end{equation}
We compute the physical $\varphi$-velocity directly from the simulation as
\begin{equation}
    v_\varphi=\frac{u^{\phi}}{u^{t}} \sqrt{g_{\phi \phi}}
    \label{eq:v-phi},
\end{equation}
\noindent
where $g$ is the metric tensor. For a given quantity $\mathcal{Q}$, we compute the mass-weighted averages over two coordinates using
\begin{equation}
    \mathcal{Q}_{\textup{avg}}=\frac{\int{\mathcal{Q} \rho u^t {\rm d}A_{\mu\nu}}}{\int{\rho u^t {\rm d}A_{\mu\nu}}},
    \label{eq:1davg}
\end{equation}
where
\begin{equation}
    {\rm d}A_{\mu\nu}=\sqrt{- g}{\rm d}\mu {\rm d}\nu.
    \label{eq:dA}
\end{equation}
\noindent
where $g$ is the determinant of the metric tensor. Radial profiles have ${\rm d}A_{\mu\nu} \propto {\rm d}\theta {\rm d}\phi$ and polar profiles have ${\rm d}A_{\mu\nu} \propto {\rm d}r d\phi$. In calculations involving radial averages, we restrict the region of integration radially to avoid capturing high-density material from the half of the star which escapes the black hole. The region of integration for a given calculation is described in more detail in the caption of the corresponding figure.

Figure \ref{fig:radial} shows that the radial profiles (i.e., averaged over angles) of density and pressure closely follow power law relationships between the inner and outer boundaries of the disk (10--200 $R_g$), hinting at a possible analytic description (see Section \ref{sec:ZEBRA}).
The angle-averaged $\varphi$-velocity is fitted by $v_\varphi \simeq 0.76r^{-0.5}$, which implies a sub-Keplerian velocity distribution that may be due to thermal pressure support against gravity (see Section \ref{sec:force}).

The internal energy density and mass density are floored at $2.27\times10^{-12}$ (see Section~\ref{sec:params}). These floors are responsible for the flat density and pressure regions at large radii in Figure \ref{fig:radial}. While these floors would have a negligible effect on a TDE at peak fallback rate, they become significant for the early times and low fallback rates considered in our simulation. The floors may affect our results by providing external pressure support to the outer disk, artificially lowering its radial and vertical extent. We discuss this in Section~\ref{sec:ZEBRA}.

Figure \ref{fig:polar} shows the polar profiles of density, pressure, and squared specific angular momentum within the disk. We calculate the pressure as above (Equation \ref{eq:pressure}) and the specific angular momentum as $l=u_\phi$. The polar profiles of all three quantities are fit to a power $\alpha$ of $\sin^2 \theta$ near the equatorial plane. We analyze these relationships further and compare them to model predictions in Section \ref{sec:ZEBRA}.

\begin{figure}
	\includegraphics[width=1.0\columnwidth]{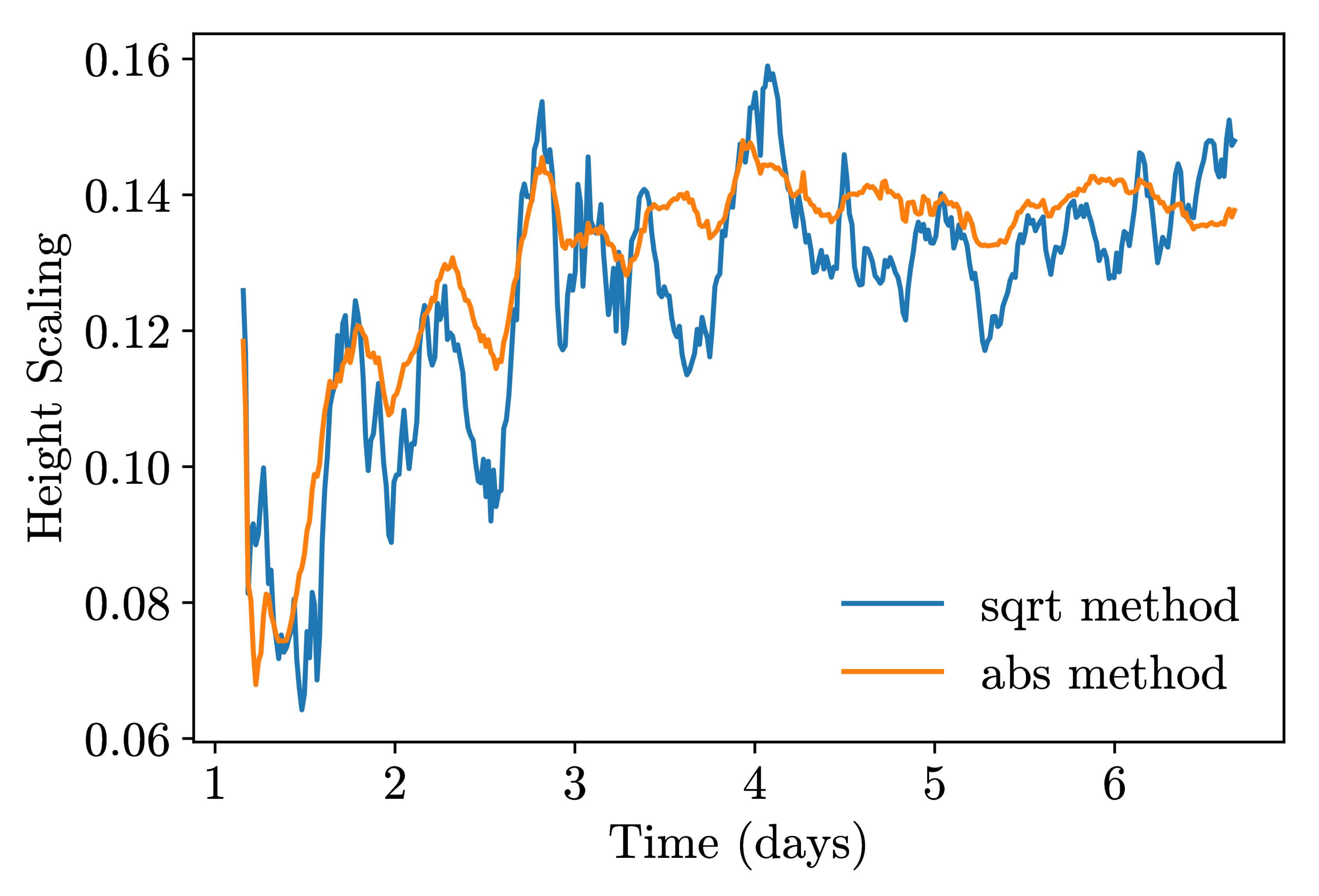}
    \caption{The scale height of the disk plotted with respect to time in our aligned TDET0 model. Scale height is calculated by averaging the mass-weighted angle from the equatorial plane over $|\theta-\pi/2|<0.3$ and $\varphi$ using two methods described by Equations \ref{eq:HoRsqrt} and \ref{eq:HoRabs}. The stream is ignored using the entropy condition. The generation of thermal energy due to stream-disk interactions and self-intersection shocks causes the gas in the disk to expand over time, increasing the scale-height.}
    \label{fig:HoR}
\end{figure}

The vertical height of the accretion disk above the midplane is proportional to the distance from the center of the BH, where the constant of proportionality is known as the scale height. We compute the scale height in 2 ways (Figure \ref{fig:HoR}).

\begin{align}
    \frac{h_{\textup{sqrt}}}{r}=\ &\sqrt{\frac{\int{ (\theta-\pi/2)^2\rho u^t dV}}{\int{\rho u^t {\rm d}V}}},
    \label{eq:HoRsqrt}\\
    \frac{h_\textup{abs}}{r}=\ &\frac{\int{ |\theta-\pi/2| \rho u^t dV}}{\int{\rho u^t {\rm d}V}},
    \label{eq:HoRabs}
\end{align}
\noindent
where
\begin{equation}
    {\rm d}V=\sqrt{-g}{\rm d}r{\rm d}\theta {\rm d}\phi
    \label{eq:dV}
\end{equation}

Both methods show that $h/r$ increases over time, indicating that the disc ``puffs up'' from the midplane. Although it is possible that the vertical expansion of the disk is artificially slowed by the pressure and density floors, this effect should not significantly impact this general trend. The increase in the angular extent of the material is due to excess thermal energy generated in the disk by the dissipation of orbital energy. The scale height reaches a plateau around the time that the self-intersections stop (3.68 days), suggesting that the violent self-intersections play a crucial role in the early heating of the disk. We discuss the mechanisms of energy dissipation further in Section \ref{sec:energy}.

\subsection{Tilted Disk Formation and Evolution}
\label{sec:tilted}

The majority of TDE disk formation simulations use either Newtonian gravity or a general relativistic treatment (exact or approximate) of a non-spinning Schwarzschild BH. However, tidally disrupted stars approach the BH from a quasi-isotropic distribution of inclinations, highlighting the importance of more general disk formation simulations that account not just for BH spin, but also for spin-orbit misalignment. Various effects unique to tilted accretion disks, such as global precession, Bardeen-Petterson alignment, and disk tearing \citep{nixon12, Liska2019b, Hawley2019} may all manifest themselves in TDE accretion disks. Although previous studies have considered tilted TDEs analytically \citep{Stone2012, Zanazzi2019}, only two numerical efforts have, to date, simulated the formation of an accretion flow following the disruption of a star on a misaligned orbit: the early work of \citet{Hayasaki2016}, and the more recent simulations of \citet{Liptai+19}. In both cases, the authors find that for adiabatic gas equations of state, it is challenging for nodal precession to cause significant delays in self-intersection. However, both of these simulations employed unrealistically eccentric stellar trajectories for computational convenience; the work presented in this section is the first numerical simulation of tilted TDEs with realistic astrophysical parameters.

\begin{figure}
	\includegraphics[width=1.0\columnwidth]{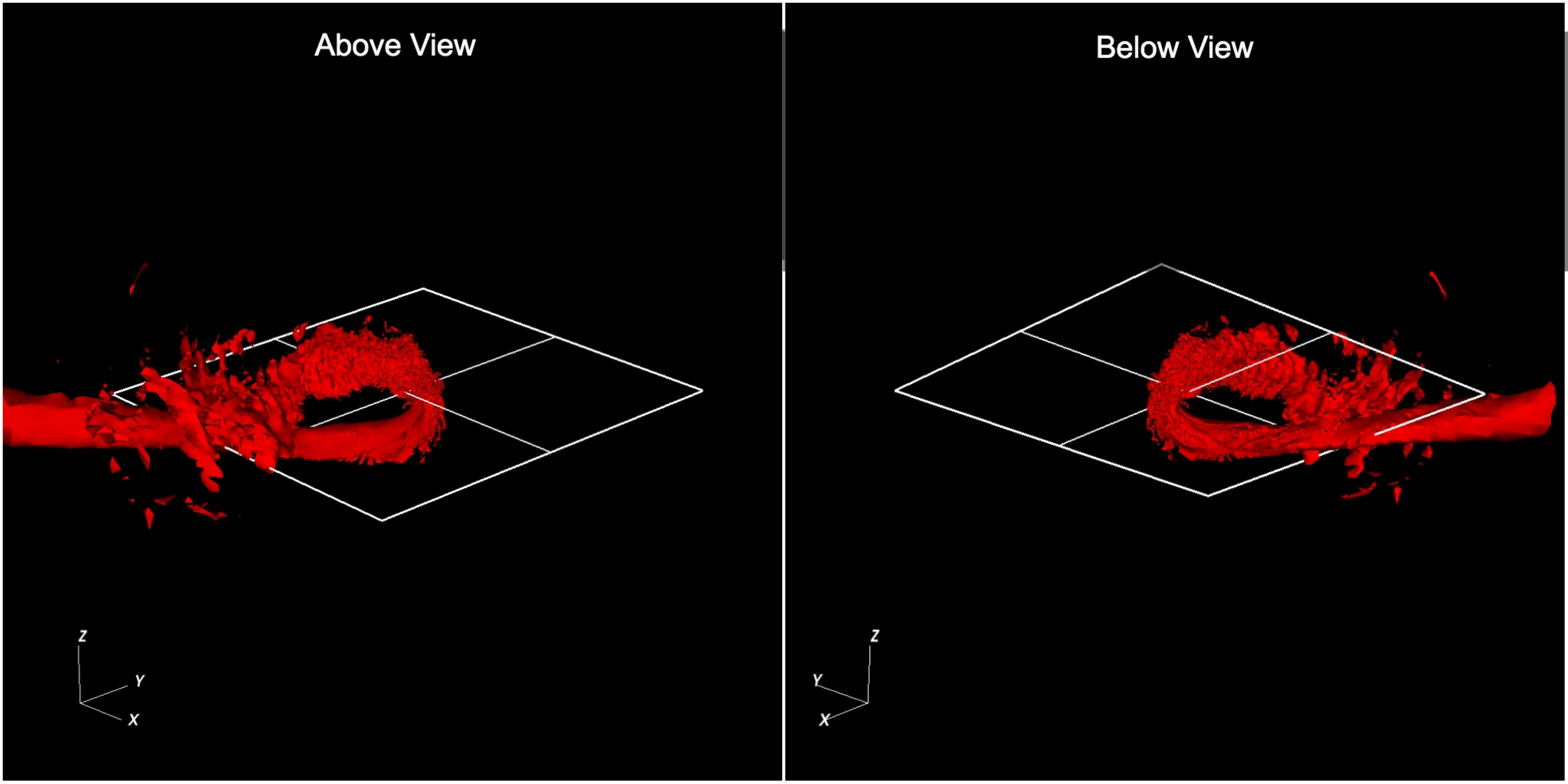}
    \caption{
    A 3D contour of density ($\rho=10^{-8}$) visualized at 3.7 days in our tilted TDET30 model. We show the views from above and below the BH orbital plane in the left and right panels, respectively. The outgoing and incoming streams are misaligned at the self-intersection point due to the nodal precession at the pericenter passage. As a result, material is ejected out of the orbital plane of the star. The orbital plane of the BH is shown for reference. To see this effect in the full context of disk formation, see the 3D renderings linked in \ref{sec:support}.}
    \label{fig:splash}
\end{figure}

The movie and 3D rendering (Section \ref{sec:support}) of the tilted TDE shows that, unlike in the aligned TDE, the returning stream is never completely interrupted by self-intersections. The misalignment between the orbital plane of the stream and the rotational plane of the BH leads to strong nodal precession upon pericenter passage. The outgoing stream exits the BH in a separate plane from the incoming stream, so when the two streams collide, they are misaligned. Figure \ref{fig:splash} shows that this misalignment launches material from both streams out of their original planes. 

From a comparison of the polar density profiles of the aligned and tilted TDE (Figures \ref{fig:polar} and \ref{fig:transverse_tilted}), the accretion disk appears significantly thicker in the tilted simulation. However, there are two caveats to this interpretation of the data. First, the launching of stream material into different orbital planes as shown in Figure \ref{fig:splash} creates pockets of high density material at large polar angles which artificially increases our estimates of disk thickness. Second, the disk material may lie in multiple orbital planes because debris which falls back at early times has more time to undergo nodal precession than debris which falls back at late times. Any dependence of the inclination angle on $r$ or $\phi$ would artificially increase our estimate of disk thickness. Additionally, the angular momenta of the gas may not have enough time to homogenize because the simulation was not run for multiple viscous times of the disk.

\begin{figure}
	\includegraphics[width=1.0\columnwidth]{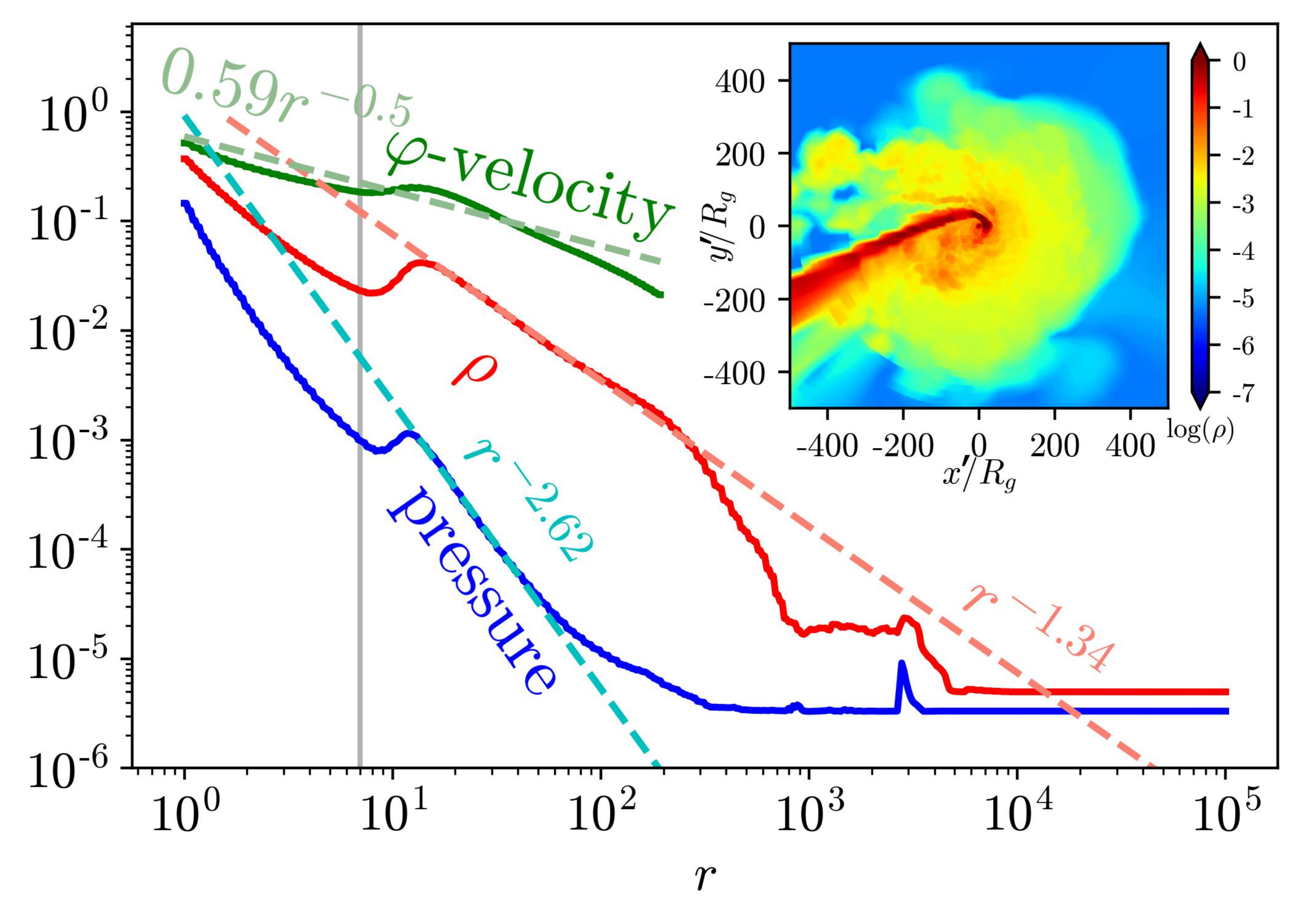}
    \caption{Analogous to Figure \ref{fig:radial}, but for the tilted TDE simulation, model TDET30. The data is tilted using our tilting algorithm (Appendix \ref{sec:tilting-algorithm}) such that the star's orbital plane coincides with $\theta'=\pi/2$. Coordinates in the tilted frame are denoted with a prime. The inset plot shows the rest mass density in the star's orbital plane at 4.0 days. See Table \ref{tab:curve_fit_tilted} for more details about the power law fits (dashed lines). Mass density and pressure are multiplied by $5 \times 10^5$ so that all three variables are roughly the same order for comparison purposes.}
    \label{fig:radial_tilted}
\end{figure}

\begin{figure}
	\includegraphics[width=1.0\columnwidth]{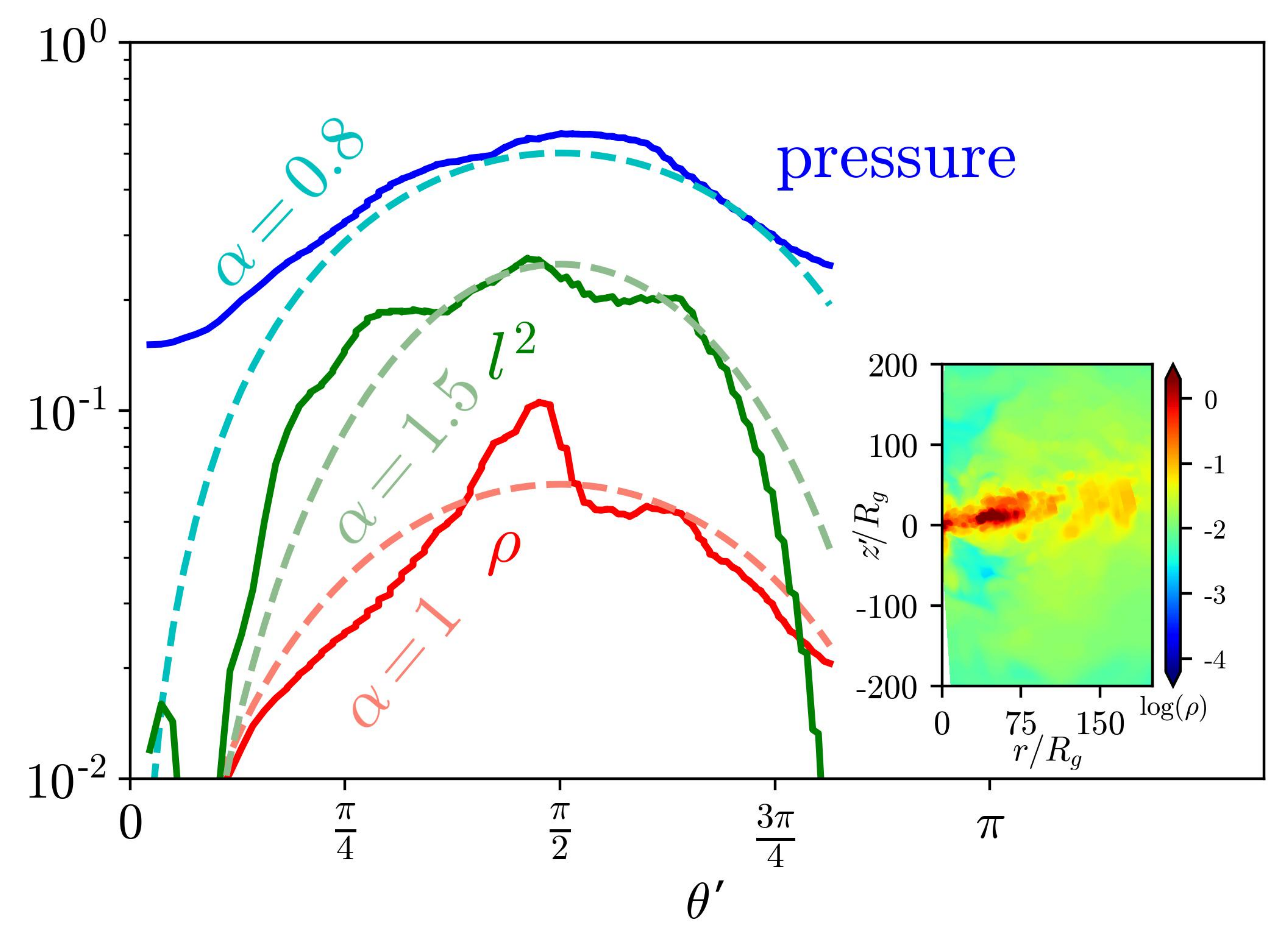}
    \caption{Analogous to Figure \ref{fig:polar}, but for the tilted TDE simulation, model TDET30. The data is tilted using our tilting algorithm (Appendix \ref{sec:tilting-algorithm}) such that the star's orbital plane coincides with $\theta'=\pi/2$. Coordinates in the tilted frame are denoted with a prime. Density and pressure are multiplied by $10^{7.3}$ and $10^{10.7}$ respectively so that all three quantities are roughly the same order for comparison purposes. The tilt angle varies over radius and time (Figure \ref{fig:angles}), so the averages may not accurately reflect the thickness of the disk.}
    \label{fig:transverse_tilted}
\end{figure}
  
Figure \ref{fig:radial_tilted} shows that the radial profiles of the tilted disk follow similar power-law relationships to those of the aligned disk, with the density and pressure falling off slightly faster in the aligned disk. As a result, the thermal pressure gradient forces in the tilted disk are larger than in the aligned disk, so the velocity distribution is more sub-Keplerian.

\section{Discussion}
\label{sec:discussion}

\subsection{Energy Dissipation}
\label{sec:energy}

The largest uncertainty in TDE evolution concerns the rate, location, and physical mechanisms that dissipate the orbital energy of dynamically cold debris streams. Past analytic models and numerical simulations examining early stages of a TDE generally focus on shock dissipation at different locations. The most important shock loci seen or proposed in past work are (i) compression shocks produced by the vertical collapse of the returning debris stream, located at the vertical caustics near pericenter \citep{Guillochon2014, Shiokawa2015}, (ii) self-intersection shocks produced at larger radii where an outgoing debris stream impacts an incoming one \citep{Rees1988, Hayasaki2013, Dai2015, Hayasaki2016}, and (iii) ``secondary shocks'' seen in the simulations of \citet{Bonnerot2019, Bonnerot2021b} after the formation of an extended accretion flow.  Each of these categories of shock have been seen to be the dominant energy dissipation mechanism at {\it some} times in {\it some} past numerical simulations of tidal disruption; however, the relative importance of these shocks is strongly affected by system parameters (e.g. SMBH mass, stellar eccentricity). The lack of published first-principles TDE simulations with astrophysically realistic parameters renders the relative importance of these shocks unclear in typical TDEs.

\begin{figure*}
	\includegraphics[width=0.8\textwidth]{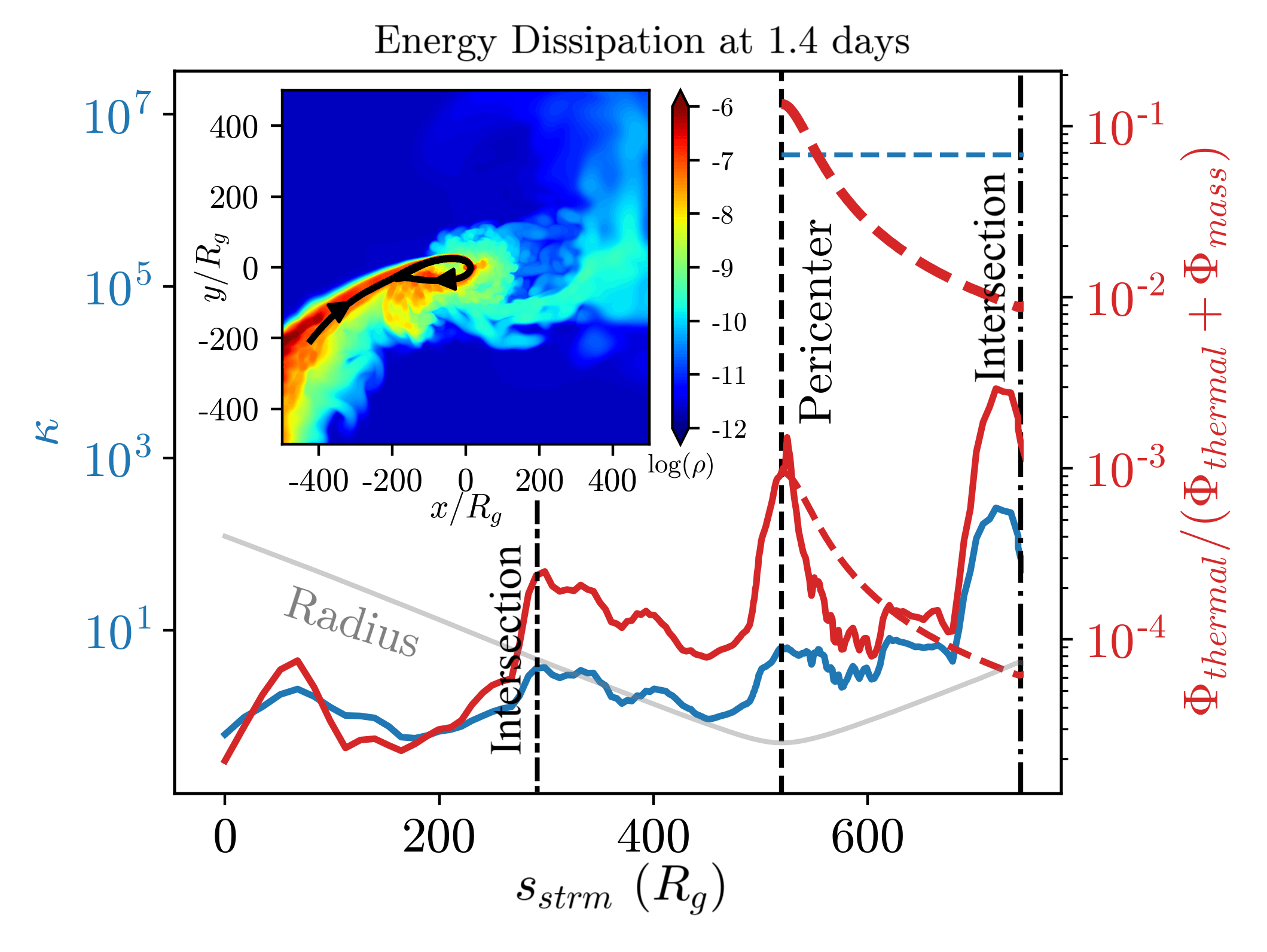}
    \caption{At early times in our aligned TDET0 simulation, the stream heats up at the pericenter and the self-intersection. However, the majority of the entropy generation occurs at the self-intersection, suggesting that the pericenter heating is nearly adiabatic. To see this, we plot the relative amount of heating $\psi$ (Equation \ref{eq:ratio}) and a proxy for entropy $\kappa \propto e^{\textup{entropy}}$ against the distance along a streamline $s_{\rm strm}$ at an early time, 1.34 days. The streamline, depicted in the inset plot, is integrated from the velocity field using a second-order Runge-Kutta method. The location of the pericenter and self-intersections along the streamline are depicted by vertical dashed and dot-dashed lines respectively. The thin gray line shows the distance between the streamline and the BH. The average entropy of the disk is shown by the blue horizontal dotted line. The thin and thick red lines show an approximation for the relative amount of heating $\psi$ for a thin ($h/r=0.082$) and thick ($h/r=1$) disk respectively (Equation \ref{eq:approx}). Note that the returning gas will not necessarily follow the path of the streamline, so $s_{\rm strm}$ is not a perfect proxy for time. For instance, the entropy sometimes decreases mildly as $s_{\rm strm}$ increases.}
    \label{fig:diss_early}
\end{figure*}

\begin{figure*}
	\includegraphics[width=0.8\textwidth]{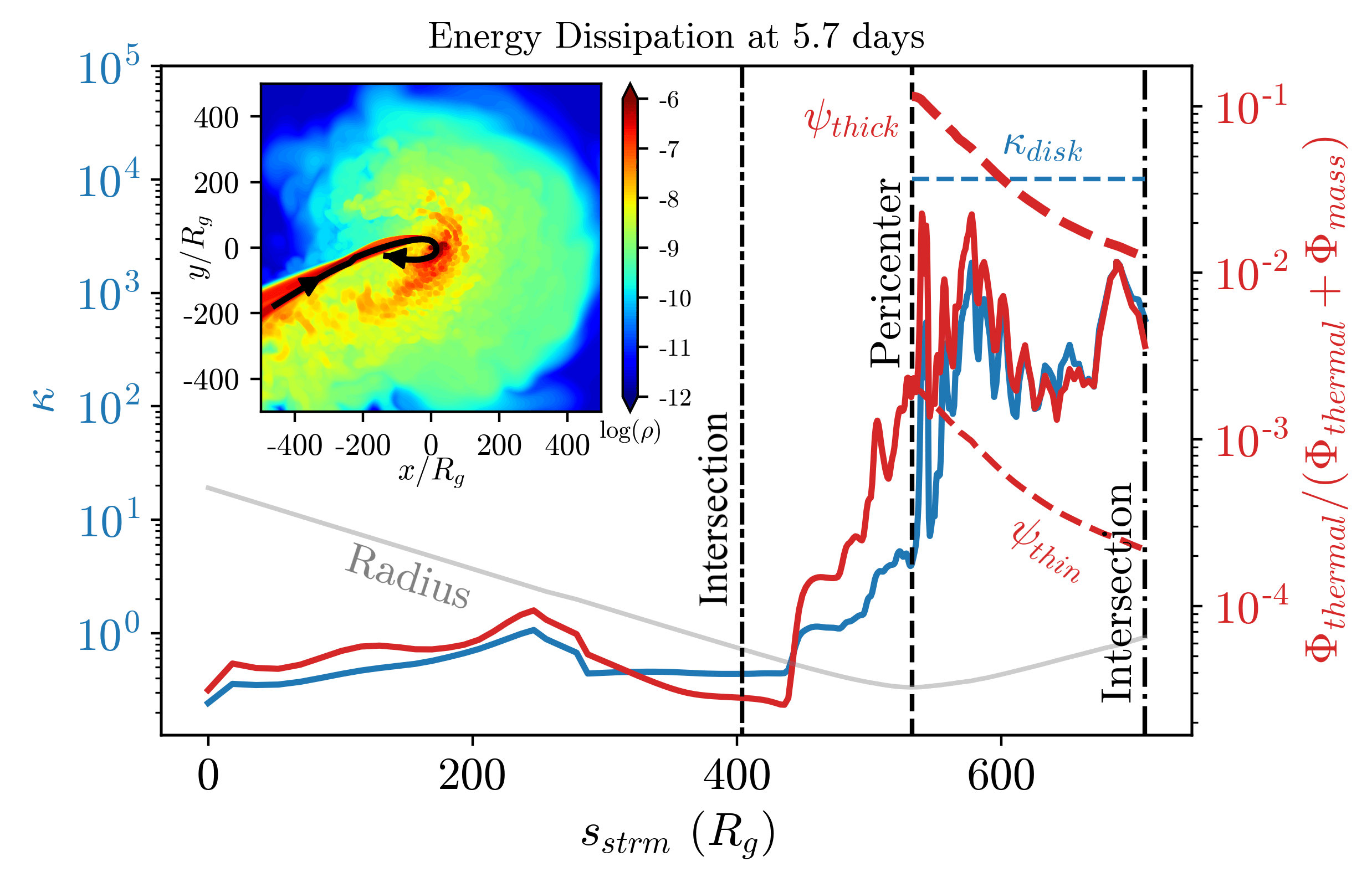}
    \caption{Analogous to Figure \ref{fig:diss_early}, but at a late time of 5.66 days. At late times in our aligned TDET0 simulation, the bulk of the heating and entropy generation occurs at the pericenter radius, suggesting that the pericenter is the most significant source of energy dissipation. The thin and thick red lines show an approximation for the relative amount of heating $\psi$ for a thin ($h/r=0.132$) and thick ($h/r=1$) disk respectively (Equation \ref{eq:approx}).}
    \label{fig:diss}
\end{figure*}

We analyze the energy dissipation of the system at both early and late times by tracking entropy and the ratio of the thermal to total energy flux along a streamline. The thermal and mass energy fluxes are given by
\begin{align}
    \Phi_{\textup{thermal}}=\ &-\sqrt{-g} (u_g + p) \eta u_t=-\sqrt{-g} u_g \gamma \eta u_t
    \label{eq:therm_flux}\\
    \Phi_{\textup{mass}}=\ &-\sqrt{-g} \rho c^2 \eta u_t
    \label{eq:mflux2}
\end{align}
\noindent
where $\eta=\sqrt{v_{\rm i} v^{\rm i}}=\sqrt{g_{\rm ij}v^{\rm i}v^{\rm j}}$ is the magnitude of the 3-velocity (we are adopting the convention where Latin indices range from 1 -- 3). We define the quantity
\begin{equation}
    \psi=\frac{\Phi_\textup{thermal}}{\Phi_\textup{thermal} + \Phi_\textup{mass}}.
    \label{eq:ratio}
\end{equation}
\noindent
to characterize the generation of thermal energy in the disk (solid red lines in Figures \ref{fig:diss_early} and \ref{fig:diss}). When thermal energy flux dominates, $\psi$ approaches unity and when the mass energy flux dominates, $\psi$ approaches zero. Shocks convert orbital energy into thermal energy, so $\psi$ increases across the self-intersection shocks. We compare the entropy of the stream post-pericenter to the average entropy of the disk, with the latter defined as
\begin{equation}
    \kappa_\textup{disk}=\frac{\int{\kappa \rho u^t dV}}{\int{\rho u^t dV}}.
    \label{eq:Sdisk}
\end{equation}
\noindent
Here $\kappa$ is defined by Equation \ref{eq:kappa} and the region of integration is defined using a stricter version of the entropy condition ($\kappa>100$) to ensure that none of the high-density, low-entropy stream material contributes to the average.

Figures~\ref{fig:diss_early} and \ref{fig:diss} show the early- and late-time dissipation profiles along a streamline, respectively. Although heating and entropy generation occur on similar levels at both times, the dissipation mechanisms are distinct. At early times, there is comparable heating as the stream passes through pericenter and as the outgoing stream reaches the intersection point. However, significant entropy generation occurs only at the self-intersection, suggesting that the heating at pericenter is nearly adiabatic while the heating at self-intersection is irreversible and shock-induced. The nearly adiabatic heating at pericenter implies that the nozzle shock is inefficient at dissipating orbital energy. 

This finding differs from other numerical work which finds that the nozzle shock is associated with a large jump in entropy \citep{Guillochon2014, Bonnerot2021}. The strong nozzle shock seen in these works requires a highly supersonic vertical collapse of the debris stream. However, we find that the vertical collapse of the debris stream at early times is subsonic, possibly due to the relatively short return time of the debris or additional heating produced in the deeply penetrating $\beta = 7$ encounter we consider. The average mass-weighted vertical Mach number in the stream at early times (distinguished by the entropy condition) is $M_z \approx 0.4$. The vertical Mach number is estimated as $M_z \approx v^z / c_s$, where $v^z$ is the magnitude of the vertical velocity and $c_s$ is the sound speed given by
\begin{equation}
    c_s^2 \approx \left( \frac{\partial p}{\partial \rho} \right)_{\mathcal{S}} = \frac{\gamma (\gamma - 1) u_g}{\rho}
    \label{eq:sound}
\end{equation}

%\DIFadd{Another effect which acts to reduce the impact of the nozzle shock is desynchronization, where material at different vertical distances within the debris stream reaches the equatorial plane at different locations along the trajectory rather than at a single point }\citep{Stone2013}\DIFadd{. Desynchronization results from a non-homologous compression of the debris stream as it approaches pericenter, which may result from the nonlinear hydrodynamics of the disruption or from fluid turbulence in the debris stream. By looking at the vertical velocity profile within vertical columns of the stream, we confirmed that the collapse is nearly homologous, with slight non-linearities arising close to the nozzle. Therefore, it is unlikely that desynchronization is a significant factor reducing the impact of the nozzle shock in our simulation.}

%\DIFadd{Finally, it is possible that we fail to capture the nozzle shock completely due to insufficient vertical resolution across the vertical extent of the stream. We leave it to future work to investigate the resolution requirements for accurately capturing the gas dynamics at the nozzle.}

At late times, the bulk of the heating and the entropy generation occurs at the pericenter. After the pericenter passage, entropy increases to more than three quarters the entropy of the disk, with the remainder of the entropy generation occurring between the pericenter and self-intersection radii.

\begin{figure}
	\includegraphics[width=1.0\columnwidth]{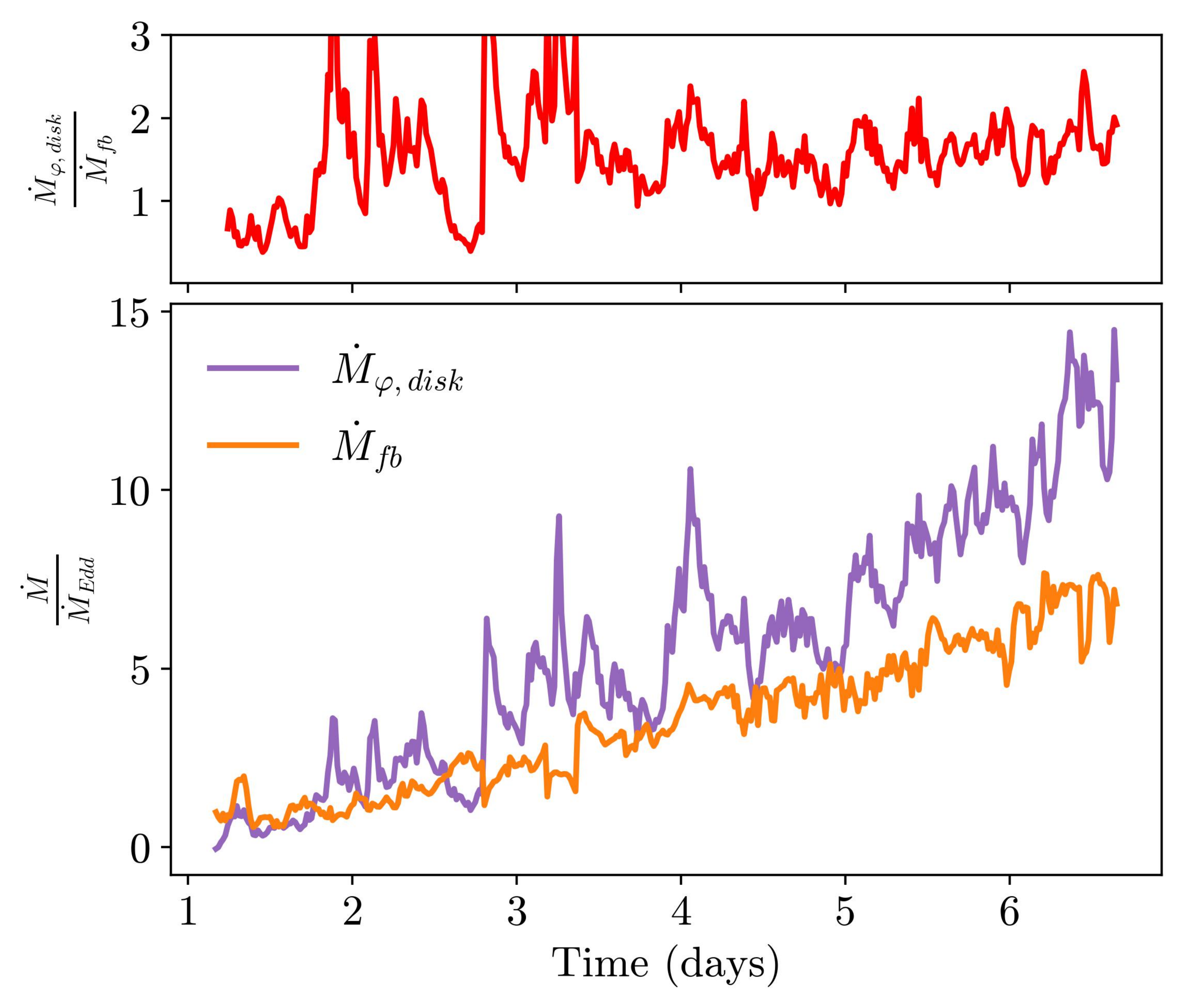}
    \caption{The mass fallback rate and the rotational mass flux in the disk in our aligned TDE simulation. The mass fallback rate is the mass flux in the stream (distinguished by the entropy condition) through the tidal sphere. The rotational mass flux is estimated by azimuthal mass flux of the disk through the surface given by $\varphi=1.14\pi$, $10<r<500$, and $\pi/2 - h/r<\theta<\pi/2+h/r$. The ratio of rotational mass flux to fallback rate is shown in the top panel. As the disk becomes denser and more massive over the course of the simulation, the rotational mass flux of the disk surpasses the mass fallback rate. Around 3.7 days, the disk completely intercepts the momentum of the outgoing stream and self-intersections can no longer occur.}
    \label{fig:diskflow}
\end{figure}

The greater value of $\psi$ along the streamline in Figure \ref{fig:diss} compared to Figure \ref{fig:diss_early} indicates that more heating takes place along the streamline at late times, possibly due to the stream-disk interactions discussed in Section \ref{sec:energy}. However, this does not imply that more heating occurs at late times in the disk as a whole. In particular, we find that the average value of $\psi$ within the disk is greater at time of Figure \ref{fig:diss_early} than at the time of Figure \ref{fig:diss}, suggesting that more relative heating occurs within the disk at early times.

As discussed in Section \ref{sec:props}, the accretion disk becomes thicker and more massive over time. Therefore, as time progresses, the disk absorbs a greater proportion of the momentum of the outgoing stream, reducing the impact of the self-intersections. When the rotational momentum of the disk surpasses the momentum of the outgoing stream, the outgoing stream disintegrates and the self-intersections stop all together.

In Figure \ref{fig:diskflow}, we plot the rotational mass flux of the disk and the mass fallback rate. The mass fallback rate is a good approximation for the mass flow rate in the outgoing stream; particularly at early times when the incoming and outgoing stream mass flow rates are most similar. At early times, the ratio of the rotational mass flux to the mass fallback rate is less than unity (Figure \ref{fig:diskflow}, top panel). This reflects that the accretion disk mass is small relative to that of the outgoing stream. The ratio of rotational mass flux to mass fallback rate only exceeds unity shortly after the first self-intersection event at 1.5 days. With each self-intersection, the disk grows more substantive until the disk completely intercepts the outgoing stream and self-intersections can no longer occur around 3.7 days in our simulation.

At early times, the incoming stream heats up before the pericenter passage at the self-intersection point. This is reflected in Figure \ref{fig:diss_early} by the dramatic increase in $\psi$ in the incoming stream at the intersection point. At the late times, this intersection is too weak to appreciably heat the incoming stream. Instead, the heating before the pericenter passage is caused by the collision between the incoming stream and the dense inner accretion disk. This is reflected in Figure \ref{fig:diss} by the increase in $\psi$ and entropy after the self-intersection point and before the pericenter passage. Together, these results suggest that the self-intersections play a larger role in the energy dissipation at early times in the TDE evolution.

To provide more context for the meaning of the quantity $\psi$, we compute an approximation for $\psi$ in a thick ($h/r=1$) and thin disk, where the thin disk approximation assumes a scale height equal to the scale height of the disk (0.069 at 1.4 days and 0.132 at 5.7 days).
\begin{equation}
    \psi \approx \frac{u_g \gamma}{\rho} \approx c_s^2 \approx \left(\frac{h}{r}\right)^2 v_k^2 \approx \left(\frac{h}{r}\right)^2 \frac{1}{r},
    \label{eq:approx}
\end{equation}
\noindent
where $c_s$ is the sound speed in the disk. This approximation appears as a dotted line in Figures \ref{fig:diss_early} and \ref{fig:diss}. Note that Equation \ref{eq:approx} is only a good approximation for small values of $\psi$. Contrary to the approximation, we find that $\psi$ does not drop off with radius, especially at late times, due to the heating that occurs as the stream disintegrates into the disk.

The inclusion of a more realistic equation of state within the debris stream is expected to yield an even higher rate of dissipation due to hydrodynamical shocks (see \citet{Guillochon2014}, Section 3.3, paragraph 6). Additionally, the inclusion of magnetic fields is expected to yield extra dissipation through the action of MRI \citep{Balbus1991} when the inner and outer stream develop a strong shear in velocity near pericenter.

\subsection{Circularization}
\label{sec:Circularization}

In the standard TDE picture, the accretion disk circularizes efficiently as shocks dissipate orbital energy, resulting in a nearly axis-symmetric accretion disk with low eccentricity. However, not all TDE disks circularize completely \citep{Piran2015}. \citet{Cao2018} find that the optical emission lines of TDE ASASSN-14li are best modelled by an accretion disk with eccentricity $e$=0.97. Furthermore, recent analytical work on TDEs has derived eccentric disk solutions which can produce radiation consistent with the X-ray and optical luminosities of many TDE candidates \citep{Zanazzi2020}.

In our simulation, the accretion disk tends towards circularization but never fully circularizes according to the criterion suggested by \citet{Bonnerot2017}: an average eccentricity lower than 1/3. Instead, our disk reaches an average eccentricity of 0.88 at late times, where eccentricity is given by
\begin{equation}
    e=\sqrt{1 + \frac{2 \varepsilon l^2}{G^2 M_{\textup{BH}}^2}}.
    \label{eq:ecc}
\end{equation}
\noindent
Here $\varepsilon=-(u_t  +  1)$ is the total orbital energy, $l=u_\phi$ is the specific angular momentum, and the average is taken from $r=10 R_g$ to $r=400 R_g$. It is important to note that this formula only provides an upper bound on the geometric eccentricity of particle trajectories because it assumes that a given fluid element is acted on only by a Newtonian gravitational force. For example, in the presence of internal pressure support, Equation \ref{eq:ecc} gives values greater than zero for circular orbits. Despite this issue, test particle eccentricity is still a useful metric for the extent of circularization.

Due to the short duration of our simulation, we cannot confirm whether more complete circularization will occur at times after the end of our simulation. One factor that may inhibit circularization is the injection of high-eccentricity material into the disk by the returning debris stream. However, this effect will become negligible when the mass fallback declines to the point where energy and angular momentum input are negligible in analogy to the late-time behavior of the disk mass in \citet{Cannizzo1990}.

\begin{figure}
	\includegraphics[width=1.0\columnwidth]{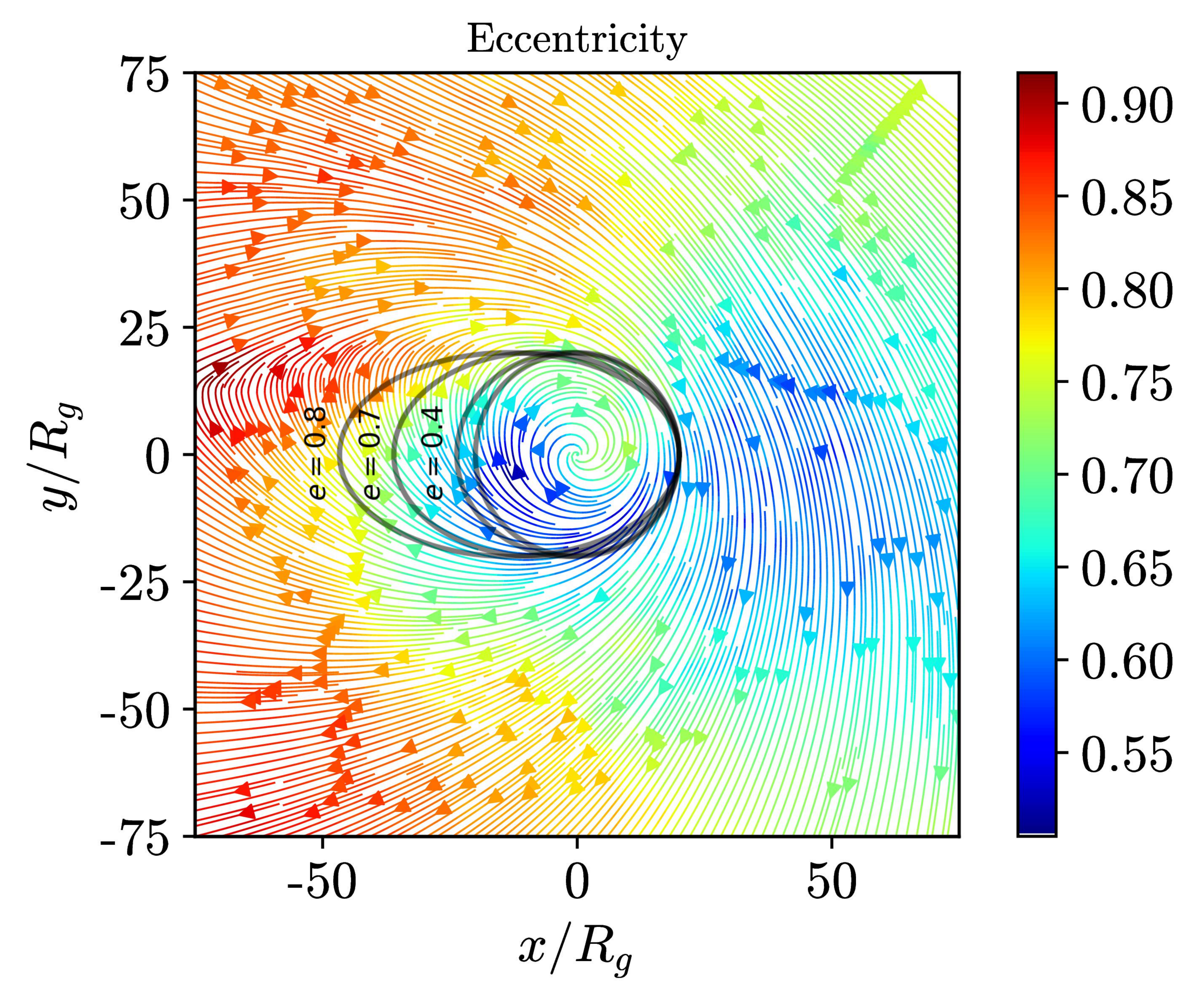}
    \caption{A time-averaged velocity streamline plot of the inner parts of the disk in the equatorial plane colored by time-averaged eccentricity. Velocity and eccentricity are time-averaged over the simulation's entire duration, averaged over $|\theta - \pi/2| < 0.5$, and weighted by rest mass density. The stream is ignored using the entropy condition. The streamlines are integrated using a second-order Runge-Kutta method. Ellipses of various eccentricities are overlaid to provide context for the eccentricity data.}
    \label{fig:Stream Plot}
\end{figure}

\begin{figure}
	\includegraphics[width=1.0\columnwidth]{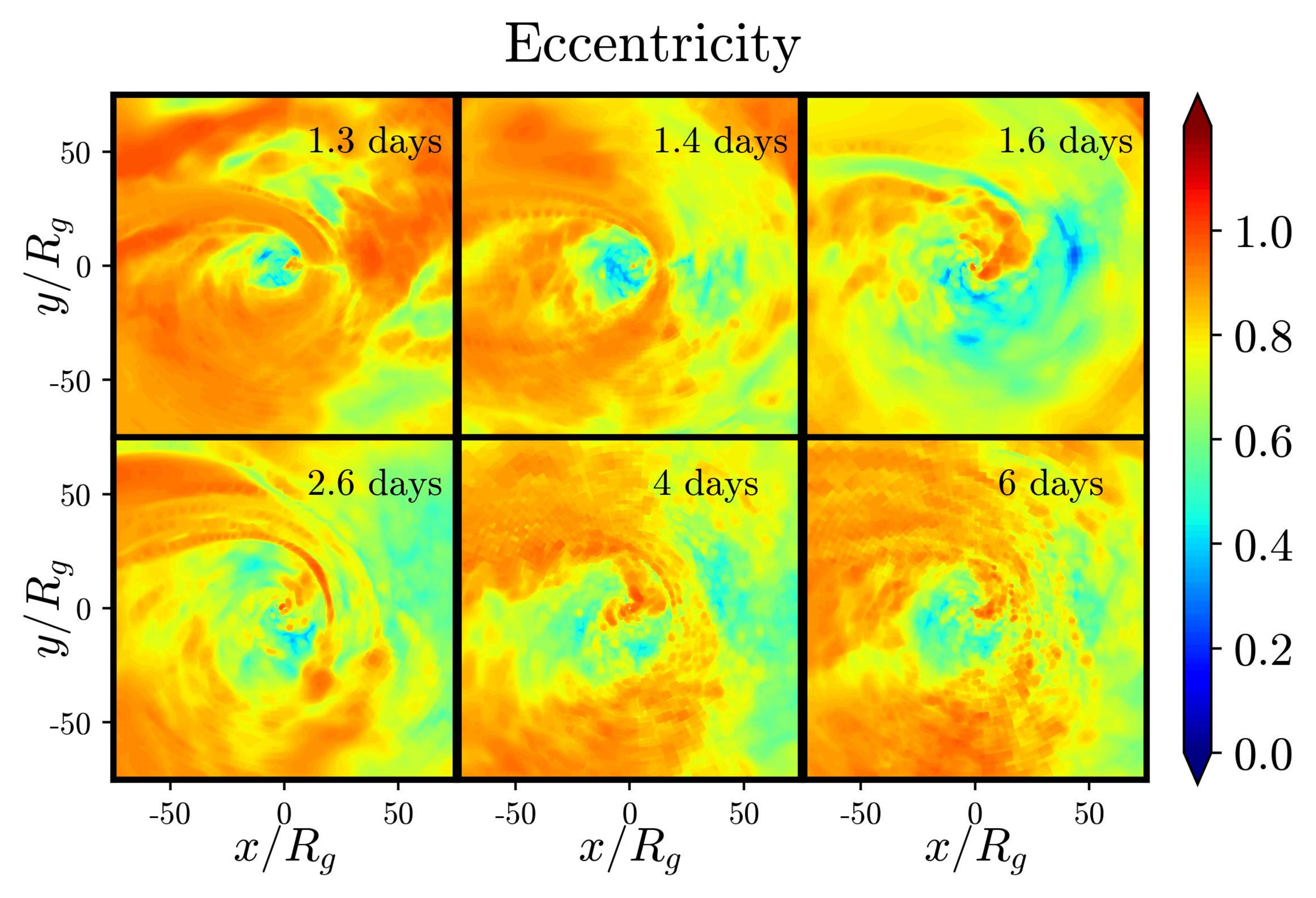}
    \caption{Snapshots of mass-weighted eccentricity in the equatorial plane at various times (see legend) averaged over $\theta = \pi / 2 \pm 0.5$. The stream is ignored using the entropy condition. However, there still is an abundance of high-eccentricity material around the stream, suggesting that the stream constantly transfers its energy and angular momentum into the disk.}
    \label{fig:Ecc xy}
\end{figure}

Figure \ref{fig:Stream Plot} shows a time-averaged streamline plot of the inner part of the disk colored by eccentricity. The area immediately surrounding the stream contains high-eccentricity material, indicating that the stream continuously transfers its energy and angular momentum into the disk. Away from the stream, disk material orbits at more moderate eccentricities. Looking at snapshots throughout the duration of the simulation, we see a similar distribution of eccentricities in the disk (Figure \ref{fig:Ecc xy}). Figure \ref{fig:Stream Plot} also shows that the eccentricity changes along each streamline, suggesting a continuous transport of energy and angular momentum within the disk itself.

\begin{figure}
	\includegraphics[width=1.0\columnwidth]{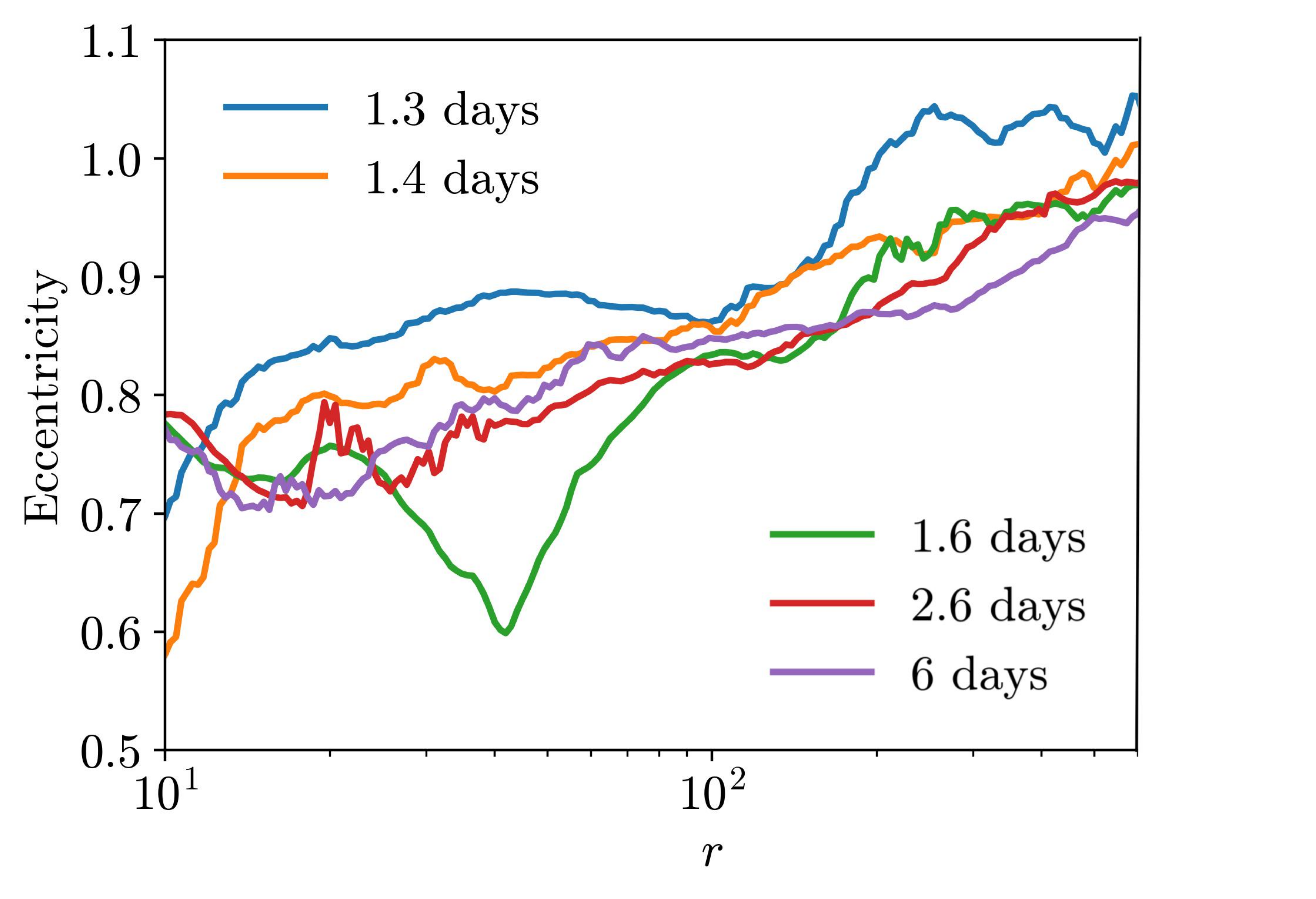}
    \caption{Radial profile of eccentricity at various times (see legend; compare to Figure \ref{fig:Ecc xy}). Eccentricity is mass-weighted and averaged over spherical shells. The stream is ignored using the entropy condition. Unbound material is ignored using the Bernoulli parameter. Only radii from $10 R_g/c$ to $500 R_g/c$ are shown. Eccentricity is unevenly distributed throughout the disk. In particular, we see greater circularization at smaller radii in the disk.}
    \label{fig:Ecc vs R}
\end{figure}

Eccentricity is not evenly distributed across the different radii in the disk (Figure \ref{fig:Ecc vs R}). In particular, the inner parts of the disk are more circularized than the outer parts. This indicates that circularization is more efficient at smaller radii, possibly because the velocity shear between neighboring radii is greater. Because eccentricity affects mean $\varphi$-velocity (Section \ref{sec:force}), the uneven eccentricity distribution may contribute to the drop in $\varphi$-velocity near the outer edge of the disk in Figure \ref{fig:radial}.

In Figure \ref{fig:Ecc vs R}, there is a dip in the eccentricity at the self-intersection radius at 1.6 days. This may be a result of the second major self-intersection event, which occurs at 1.47 days, interrupting the incoming stream. This is visible in Figure \ref{fig:Ecc xy}, where we can see that the effect of the stream on the disk eccentricity is much weaker than at any other time slice shown.

\subsection{Accretion and Outflow}
\label{sec:accretion}

\begin{figure}
	\includegraphics[width=1.0\columnwidth]{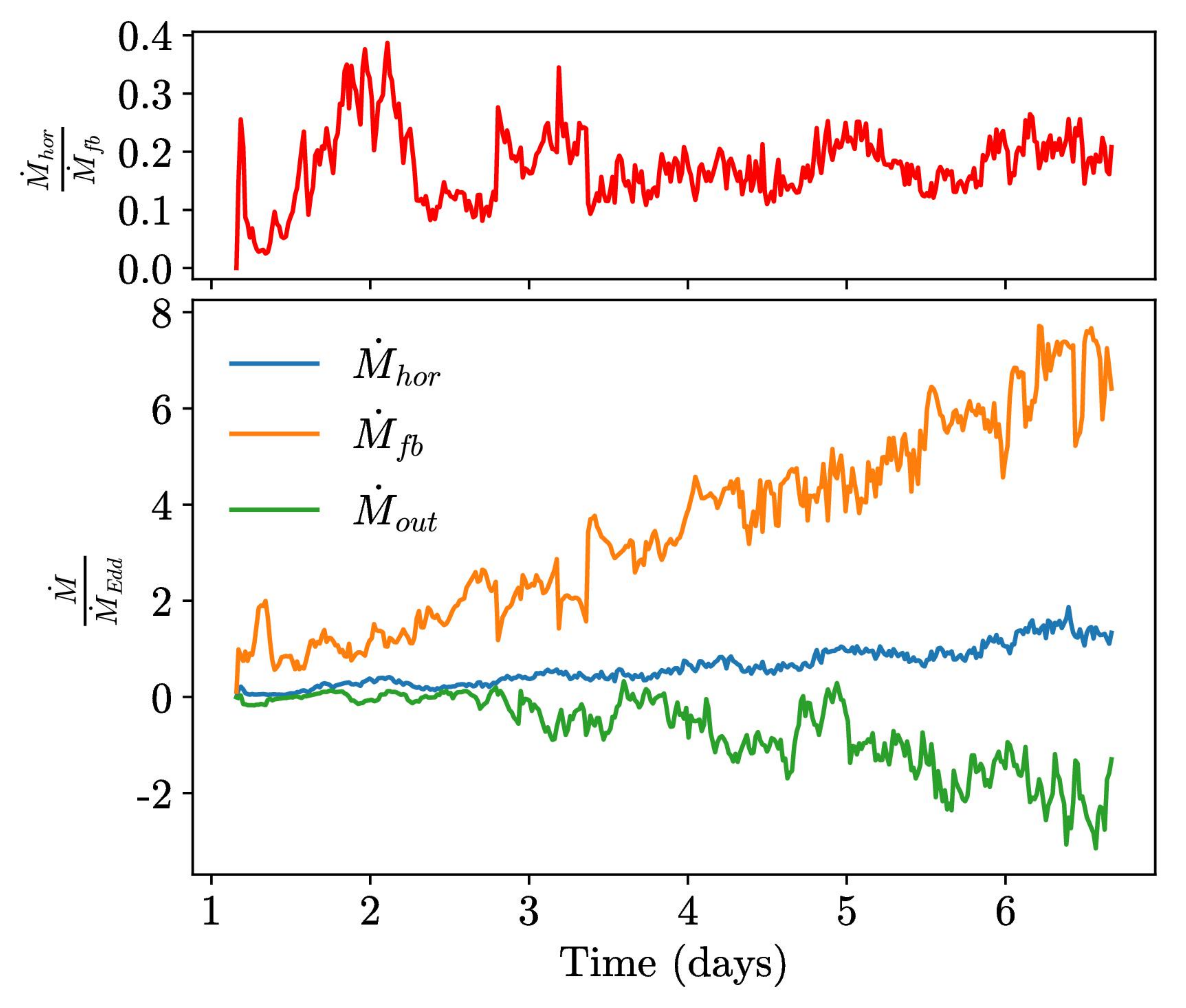}
    \caption{Mass fallback rate, mass accretion rate at the event horizon, mass outflow rate, and the accretion efficiency plotted versus time in our aligned TDET0 simulation. The accretion efficiency settles to around 10 to 20 percent at $t\gtrsim 4$ days. Positive mass fluxes are directed towards the BH. The accretion efficiency is calculated as a ratio of the mass flux at the event horizon to the mass fallback rate. Mass fallback rate is computed within the stream (distinguished by the entropy condition) through the tidal sphere. Mass outflow rate is computed as the unbound mass flux through the tidal radius. Bound matter is ignored using the Bernoulli parameter. All three mass fluxes increase roughly linearly with time, suggesting that the disk mass increases quadratically.}
    \label{fig:Accretion}
\end{figure}

Some TDE accretion models predict a period of super-Eddington accretion, the magnitude and duration of which depend on the fallback rate \citep{Coughlin2014, Wu2018}. This prediction is supported by observations; for instance, TDE Swift J1644+57 exhibits a super-Eddington luminosity \citep{Burrows2011, Zauderer2011}. We calculate the theoretical Eddington accretion rate for our simulation below. Assuming that the accreting material is mostly ionized hydrogen gas, the Eddington luminosity is
\begin{equation}
    L_{\textup{Edd}}=\frac{4 \pi G M_{\textup{BH}} c m_p}{\sigma_T},
    \label{eq:L Edd}
\end{equation}
\noindent
where $m_p$ is the mass of the proton and $\sigma_T$ is the Thomson cross section. From the Eddington luminosity, the Eddington accretion rate is
\begin{equation}
    \dot{M}_{\textup{Edd}}=\frac{L_{\textup{Edd}}}{\epsilon c^2},
    \label{eq:M Edd}
\end{equation}
\noindent
where $\epsilon$ is the gravitational potential energy that is radiated as a fraction of the rest-mass energy. Combining the above two expressions yields
\begin{equation}
    \dot{M}_{\textup{Edd}}=\frac{4 \pi G M_{\textup{BH}} m_p}{\sigma_T \epsilon c}.
    \label{eq:M Edd final}
\end{equation}

If we assume that $\epsilon = 0.1$, then we find $\dot{M}_{\textup{Edd}} \simeq 0.022 M_\odot \textup{yr}^{-1}$. Figure \ref{fig:Accretion} shows that the mass accretion rate at the BH reaches up to twice the Eddington limit, and the mass fallback rate reaches up to 8 times the Eddington limit. This confirms that the TDE in our simulation exhibits the predicted period of super-Eddington accretion. We noted in Section \ref{sec:phantom} that the maximum fallback rates in our simulation are about an order of magnitude lower than the peak fallback rate due to the short duration of our simulation. Therefore, at its peak the accretion rate would be even more super-Eddington than seen in Figure \ref{fig:Accretion}.  

\begin{figure}
	\includegraphics[width=1.0\columnwidth]{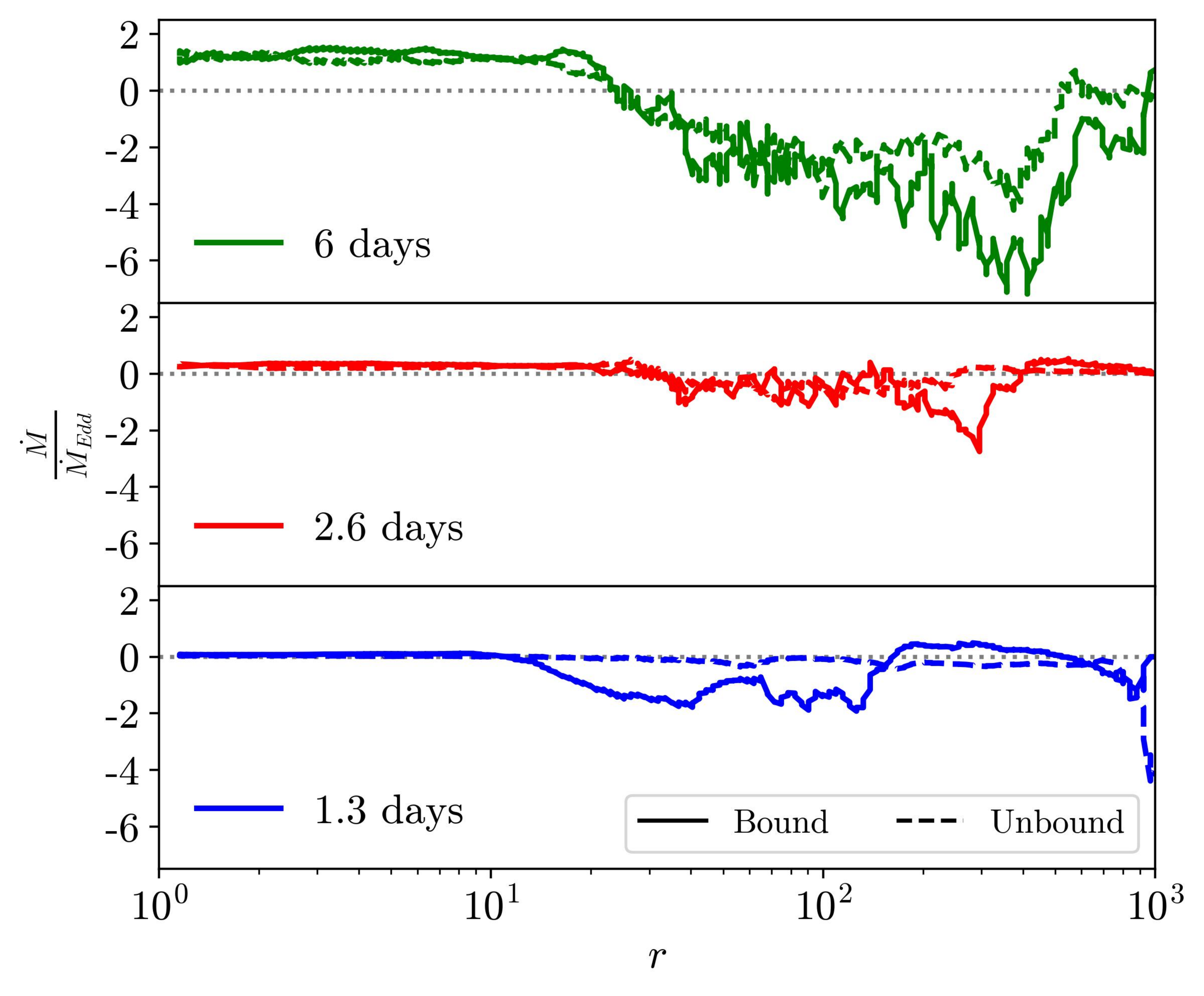}
    \caption{Radial profiles of the mass flux in the disk in units of the Eddington accretion rate. We distinguish between the bound and unbound material at three different times, spread evenly across the duration of the simulation. Positive mass fluxes are directed towards the BH. Bound and unbound material is distinguished using the Bernoulli parameter. The mass flux of bound material is shown in solid lines, and the mass flux of unbound material is shown in dotted lines. The stream is ignored using the entropy condition. The horizontal dotted lines show $\dot{M}=0$. Bound and unbound material accrete onto the BH at increasingly higher rates as the simulation progresses.}
    \label{fig:Mass Flux}
\end{figure}

Figure \ref{fig:Mass Flux} shows that bound and unbound material accrete onto the BH at increasingly higher rates as the simulation evolves. However, outside of the innermost radii in the disk, more material flows away from the BH than towards the BH. This outward mass flux drives the radial expansion of the disk. The figure also shows that the radial mass fluxes of bound and unbound material are similar throughout the simulation, indicating that a significant fraction of the initially bound material gets unbound. We can loosely estimate this fraction by comparing the outflow rate to the mass flux fallback in Figure \ref{fig:Accretion}. Due to the short duration of our simulation, we compute outflow rates as the mass flux of unbound disk material through the tidal sphere. In a longer simulation, we could compute a more precise outflow rate by computing the mass flux at larger radii and at later times. Figure \ref{fig:Accretion} shows that the outflow rate is approximately 1/3 of the fallback rate, suggesting that this same fraction of infalling material becomes unbound. However, it is possible that the outflows may be artificially suppressed by the density and pressure floors discussed in Section \ref{sec:props}, so this may be an underestimate for the fraction of material which becomes unbound in a physical TDE.

The rate at which mass is added to the disk is given approximately by subtracting the mass flux outflow and mass accretion rate from the mass flux fallback. Because all three mass fluxes increase linearly with time (Figure \ref{fig:Accretion}), the mass of the disk increases quadratically over the course of our simulation. We verify this by fitting the disk mass to a quadratic time-dependence using the least-squares method. We find that the disk mass in units of grams is given approximately by $M_{\rm disk}/M_\odot = 4.85 \times 10^{-8} t_{\rm days}^2$ with a coefficient of determination $R^2 = 0.994$.

\begin{figure*}
    \centering
    \includegraphics[width=1.0\linewidth]{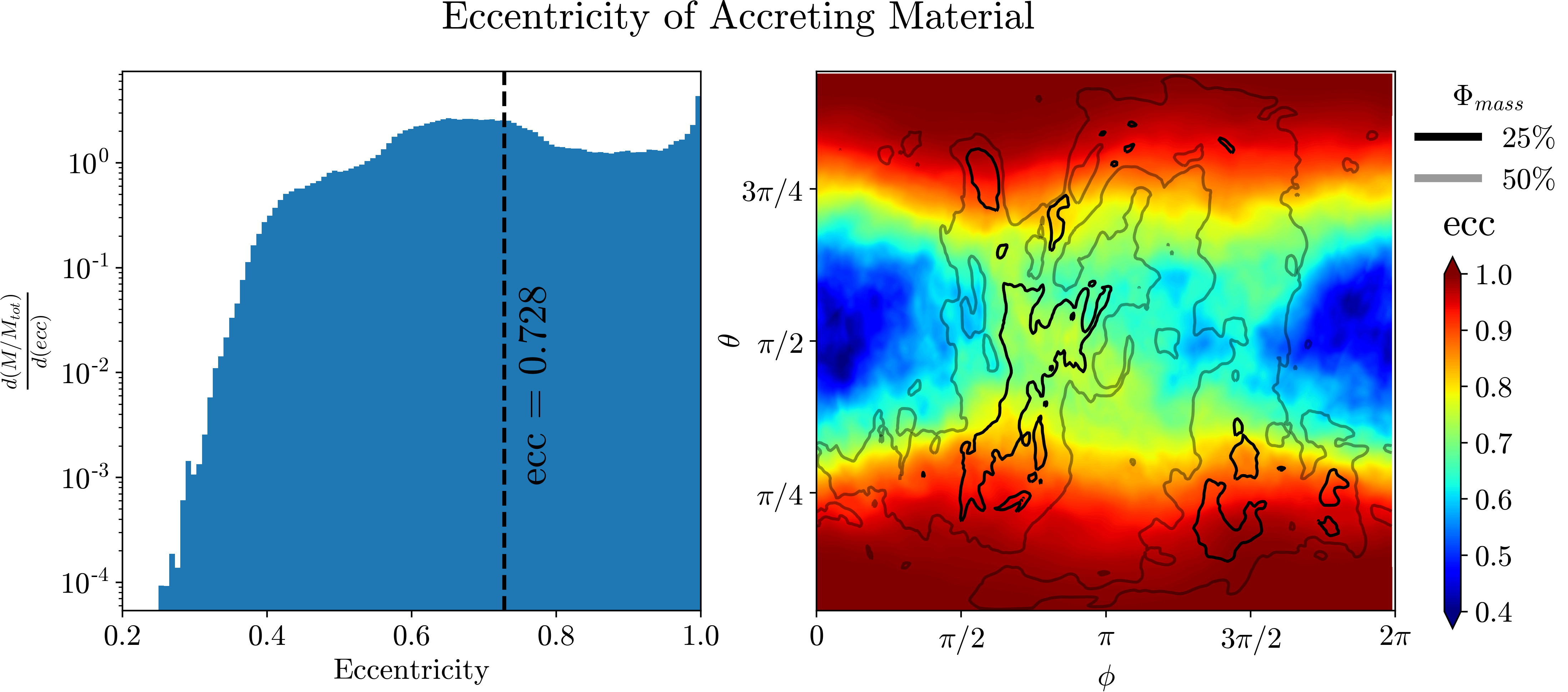}
    \caption{The left panel shows a time-averaged histogram of the eccentricity of material inside $R_{\rm ISCO}$, weighted by rest-mass. Each bin is normalized by total accreting mass and bin width. The mean eccentricity, $\langle e\rangle \approx 0.73$, is shown by the vertical dashed line. The right panel shows a time averaged colour map of the eccentricity of material inside $R_{\rm ISCO}$, weighted by rest mass. The overlaid grayscale contours outline regions containing 25 and 50 percent of the total mass flux. Floor material is ignored using an angular momentum condition ($|u_\phi| > 10^{-5}$). The left panel indicates that a significant fraction of material accretes with moderate eccentricities, allowing for a high accretion efficiency without complete circularization. However, from the right panel, we see that highly eccentric material preferentially accretes at high latitudes, except for the material at azimuthal angles just beyond that of the pericenter (see the text for discussion).  Most of the low-latitude (disk) accretion occurs at moderate eccentricity.}
    \label{fig:ecc-accretion}
\end{figure*}

We also quantify the accretion efficiency (Figure \ref{fig:Accretion}) using the ratio of the mass fallback rate to the mass accretion rate at the event horizon, where the mass flux is computed as
\begin{equation}
    \dot{M}=-\int{\rho u^r {\rm d}A_{\theta\phi}}.
    \label{eq:mflux}
\end{equation}
The accretion efficiency exhibits some periodicity due to the periodic self-intersection of the stream (Section \ref{sec:accretion}). However, after the initial spike at two days into the disruption, it settles into a range of 10 to 20 percent, suggesting that our TDE is reasonably efficient at getting the gas from the debris stream to the black hole.

If all accreting material were fully circularized, the accretion efficiency would place a lower limit on the extent of circularization. Then, if no circularization occurred, then the mass accretion rate at the event horizon would vanish, and if the disk circularized completely, then part of the material that falls back would eventually accrete (the rest would fly out as an outflow), giving the lower limit to extent of circularization.

However, we find that a significant fraction of the stellar debris accretes with moderate or high eccentricities, as shown in the left panel of Figure \ref{fig:ecc-accretion}. Here, we plot a time-averaged histogram of the eccentricity of material inside the innermost stable circular orbit, $R_{\rm ISCO}=2.04 R_{\rm g}$ (for prograde orbits in the equatorial plane and our black hole spin value of $a=0.9375$; \citealt{Bardeen1972}).

Eccentric accretion has been found in several earlier works \citep{Shiokawa2015, Sadowski2016, Bonnerot2019} and may explain the low luminosity and temperature observed in TDE candidates. A more extreme model was proposed by \citet{Svirski2017} in which magnetic stresses transport the angular momentum away from the black hole, driving eccentric accretion. However, this model does not apply to our simulation, which does not include magnetic fields.

There are two ways that material can accrete with moderate or high eccentricities. Gas in the disk may lose angular momentum through, e.g., turbulent viscosity, shocks, or spiral waves, until it plunges into the black hole. This process is analogous to accretion in a quasi-circular accretion disk. Alternatively, gas at high latitudes may get torqued and free fall directly into the black hole, an effect which has been observed in previous works (see Figure 12 of \citet{Sadowski2016}). We call the first type of accretion eccentric disk accretion and the second type of accretion ballistic polar accretion.

We can differentiate between these two types of accretion by the latitude at which matter enters the region $r<R_{\rm ISCO}$. We find that highly eccentric material ($e > 0.7$) preferentially accretes at high latitudes, suggesting that the dominant method of highly eccentric accretion is ballistic polar accretion (Figure \ref{fig:ecc-accretion}). Material accreting at low latitudes generally has more moderate eccentricities ($0.4 < e < 0.7$), and is thus likely driven by eccentric disk accretion. However, there is a patch of high eccentricity material at low latitudes in the region $2\pi/3 < \phi < \pi$, just past the pericenter at an azimuthal angle of $\phi = \pi/2$. Accretion in this region may be due to coherent chunks of the incoming stream that undergo turbulent exchange angular momentum with the disk and accrete directly onto the black hole (see also Fig.~\ref{fig:Ecc xy}). The innermost (25 percent) mass flux contour in Figure~\ref{fig:ecc-accretion} indicates that the majority of accretion occurs in this region.

As we describe in Section \ref{sec:props}, the debris stream in our aligned TDE simulation collides with itself in five violent self-intersection events that occur approximately 12 hours apart and last for roughly $2000 R_{\rm g}/c$, or 2.74 hours. We apply these results to TDE Swift J1644+57, which exhibits quasi-periodic flaring during the first few days of its initial evolution. Other authors have proposed that this flaring is due to a precessing jet \citep{Stone2012, Tchekhovskoy2013}. However, our simulations show that even without precession, the flaring due to violent self intersections can explain both the number of flares and their timescale. Swift J1644+57 is a $10^5-10^6$ $M_{\odot}$ BH, so 1 day corresponds to a timescale of 5000-50000 $R_{g}/c$, similar to the timescale of the self-intersections in our simulation.

It is unlikely that this flaring is a direct consequence of self-intersection events because the material at the self-intersection point is too optically thick to produce X-rays without adiabatic cooling \citep{Jiang2016}. Instead, we propose that the periodicity of the self-intersections leads to a periodicity in the accretion that feeds the jets, an effect that we see in our aligned TDE simulation. 

As we discuss in Section \ref{sec:Circularization}, we can normalize the mass accretion rate at the event horizon by the mass fallback rate for the aligned (Figure~\ref{fig:Accretion}) and tilted (Figure~\ref{fig:accretion_tilted}) simulations. We see quasi-periodic behavior only in the aligned case where violent, periodic self-intersections occur. This behavior does not perfectly correlate with the major self-intersection events in the simulation, which may be due to the similar timescale of the self-intersections ($\sim12$ hours apart lasting $\sim3$ each) and the fallback time from the self-intersection point ($\sim4$ hours). However, the large fluctuations in accretion rate stop after the last major self-intersection event at 3.7 days.

\citet{Sadowski2016} found a marginally bound torus after self intersection with some unbound material at high polar angles. They also found periodic behavior due to the interactions of the outgoing stream with the incoming stream. However, this interaction was not as violent as in our simulation, which may be due to the differences in the orbital properties of the initial star.

\subsection{Force Balance}
\label{sec:force}

In Section \ref{sec:props}, we show that the $\varphi$-velocities in the accretion disk are 76 percent of the expected $\varphi$-velocities in a circularized Keplerian accretion disk. There are two possible explanations for our findings. First, the non-zero eccentricity of the disk decreases the average $\varphi$-velocity of the disk relative to a completely circularized disk. Second, the disk is internally supported by non-gravitational forces.

As we discuss in Section \ref{sec:Circularization}, the average eccentricity of the disk at late times is $e = 0.88$. To determine the effect of this eccentricity on the velocity distribution, we set up artificial velocity fields with constant eccentricities of 0 and 0.88 using a method described in Appendix \ref{sec:fields}. For each velocity field, we compute the radial profile of $\varphi$-velocity with Equation \ref{eq:1davg}, where the density weight is determined as a function of radius by the power law relationship depicted by in Figure \ref{fig:radial}. On average, we find that the $\varphi$-velocities in the eccentric disk are 93.8 percent of the $\varphi$-velocities in the circularized disk. Therefore, the $\varphi$-velocities in our accretion disk are at most $76\% / 0.938 = 81\%$ of the $\varphi$-velocities in a Keplerian accretion disk at the same eccentricity. Therefore, the disk must also be externally supported by non-gravitational forces.

\begin{figure}
	\includegraphics[width=1.0\columnwidth]{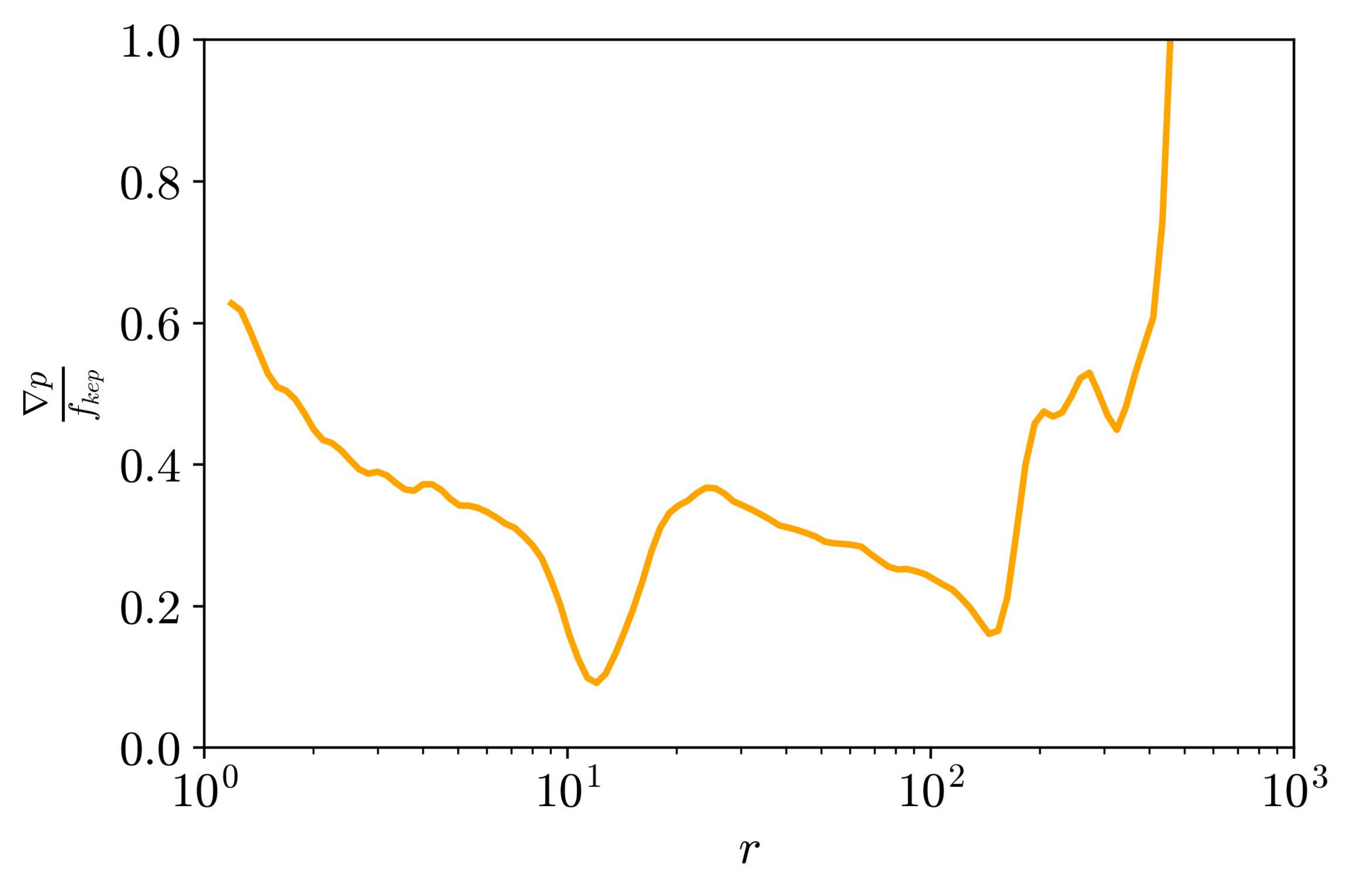}
    \caption{A time-averaged radial profile of the pressure gradient force density normalized by the Keplerian force density. Within the disk, we see that pressure gradient forces are a significant fraction of the Keplerian force, accounting for the sub-Keplerian $\varphi$-velocity distribution that we find in Figure \ref{fig:radial}. Pressure is averaged over spherical shells and mass-weighted. Time averages are over the simulation's full duration. The stream is ignored using the entropy condition. We apply a smoothing function to the data to improve readability.}
    \label{fig:Force}
\end{figure}

The only non-gravitational forces in our simulation are thermal pressure-gradient forces. Because pressure drops off as the distance from the BH increases (Figure \ref{fig:radial}), these forces are directed away from the BH in the equatorial plane, reducing the centripetal force on the accretion disk. To analyze the force balance in the disk, we compute the ratio of the pressure-gradient force density, $\nabla p$, to the centripetal force density required to maintain Keplerian orbits (Figure \ref{fig:Force}), where the Keplerian centripetal force density is

\begin{equation}
    f_\textup{kep}=\rho \frac{v_K^2}{r}=\rho \frac{GM_{\textup{BH}}}{r^2}.
    \label{eq:Force}
\end{equation} 
As the ratio of force densities increases, the matter in the disk forms stable orbits at increasingly sub-Keplerian velocities. We find that the gradient force is a substantial fraction of the Keplerian centripetal force at all radii in the disk allowing the disk material into maintain sub-Keplerian velocities. In the inner parts of the disk, the pressure-gradient force ranges from 25 to 40 percent of the Keplerian centripetal force, which fully accounts for the sub-Keplerian $\varphi$-velocity distribution.

\subsection{Comparison to Disk Models}
\label{sec:disk}

\subsubsection{Comparison to ZEBRA Model}
\label{sec:ZEBRA}

We compare the properties of the post-intersection accretion flow in our simulation with those predicted by the ZEro-BeRnoulli Accretion (ZEBRA) model proposed by \citet{Coughlin2014}. The key assumptions of the ZEBRA model are that

\begin{enumerate}
    \item the Bernoulli parameter $b$ is zero everywhere,
    \item the potential is Newtonian,
    \item the magnetic energy density is not sufficient to destabilize the disk with respect to the H{\o}iland criteria.
\end{enumerate}

\citet{Coughlin2014} show that assumption (i) ensures that the disk is gyrentropic; that is, surfaces of constant entropy, angular momentum, and Bernoulli parameter coincide. From assumption (ii), Newton's law, and the Bernoulli equation, \citet{Blandford2004} derive self-similar solutions for gyrentropic disks with an arbitrary Bernoulli parameter. The ZEBRA solutions are a special case when the Bernoulli parameter is zero everywhere.

Assumption (iii) trivially holds in our simulations due to the absence of magnetic fields. We show that assumption (i) holds in Figure \ref{fig:Bernoulli}, which depicts a histogram of the Bernoulli parameter weighted by mass in the initial conditions. We calculate that the average mass-weighted Bernoulli parameter in the initial conditions is $3.6\times 10-5$. The Bernoulli parameter in the disk is larger than this initial parameter because only the most bound debris reaches the black hole over the duration of our simulation. However, even at late times, the Bernoulli parameter is smaller than the gravitational binding energy within the disk. At 5.7 days into the simulation, the average mass-weighted Bernoulli parameter is approximately $1.5\times 10^{-3}$, or 38 percent of the binding energy in the disk at a characteristic radius $r_0 = 259 R_{\rm g}$ determined by the mass-weighted mean radial coordinate of the disk.

Assumption (ii) becomes less accurate as the radial coordinate approaches the gravitational radius. However, we find that the power law relationships predicted by the ZEBRA model extend nearly to the inner boundary of the disk as shown in Figure \ref{fig:radial}. The ZEBRA model self-similar solutions are
\begin{equation}
    \rho(r, \theta)=\rho_0\left(\frac{r}{r_0} \right)^{ - q}(\sin^2\theta)^{\alpha},
    \label{eq:Zebra1}
\end{equation}
\begin{equation}
    p(r, \theta)=\beta\frac{GM_{\textup{BH}}\rho_0}{r}\left(\frac{r}{r_0}\right)^{ - q}(\sin^2\theta)^{\alpha},
    \label{eq:Zebra2}
\end{equation}
\begin{equation}
    l^2(r,\theta)=aGM_{\textup{BH}}r\sin^2\theta.
    \label{eq:Zebra3}
\end{equation}

\noindent
These solutions describe the accretion flow density, pressure, and squared specific angular momentum, respectively, with the additional definitions

\begin{equation}
    \alpha=\frac{1 - q(\gamma - 1)}{\gamma - 1},
    \label{eq:alpha}
\end{equation}
\begin{equation}
    \beta=\frac{\gamma - 1}{1+ \gamma - q(\gamma - 1)},
    \label{eq:Beta}
\end{equation}
\begin{equation}
    a=2\frac{1 - q(\gamma - 1)}{1+ \gamma - q(\gamma - 1)},
    \label{eq:a}
\end{equation}

\noindent
where $r_0$ is some characteristic radius in the disk and $\rho_0$ is the density at that radius in the midplane. Of particular importance to our analysis are the following relationships:

\begin{enumerate}[(a)]
    \item $\rho \propto r^{-q}$,
    \item $p \propto r ^ {-q-1}$,
    \item $1/2 < q < 3$,
    \item $l \propto \sin^2\theta$,
    \item $\rho \propto (\sin^2\theta)^\alpha$,
    \item $p \propto (\sin^2\theta)^\alpha$,
    \item Equation \eqref{eq:alpha}.
\end{enumerate}

We compare the ZEBRA model predictions to the radial and polar profiles of our simulated disk (Figures \ref{fig:radial} and \ref{fig:polar}). (a) and (b) imply that density and pressure depend on $r$ as a power-law and (e) and (f) imply that density and pressure depend on $\theta$ as a power law of $\sin^2 \theta$. These predicted dependencies provide a reasonable fit for our data within the boundaries of the disk (Tables \ref{tab:curve_fit} and \ref{tab:curve_fit_tilted}). Our fitted power-law exponents for the radial profiles density and pressure differ by 1.22, which nearly matches the difference of 1.0 predicted by (a) and (b). In the model of \citet{Coughlin2014}, the power-law index $q$ can be constrained by the mass inflow rate and prescribed disk physics (e.g. the fact that angular momentum is efficiently transported in the disc), but in general it is expected to be on the order of $\sim 1-2$ (see Figure 8 of \citealt{Coughlin2014} and Figure 4 of \citealt{Wu2018}), which is exactly what we see in our simulation.

However, the alpha parameter does not match its predicted value from the ZEBRA model. The power law exponent for density indicates that $q \sim 1$. Therefore, $\alpha\sim0.5$ by equation \eqref{eq:alpha}. Instead, we find values for $\alpha$ of unity and 12 from pressure and density respectively. In addition, $l^2$ is proportional to $(\sin^2\theta)^{2.2}$ rather than $\sin^2\theta$. These discrepancies indicate that the disk must be thinner than predicted by the ZEBRA model.

The ZEBRA model predicts that the specific angular momentum of the disk must be at least 76 percent of the Keplerian value with our assumption of a polytropic index of 5/3 \citep{Coughlin2014}. Coincidentally, we find that the $\varphi$-velocities in the disk are 76 percent of the Keplerian values for circular orbits. As we discuss in Section \ref{sec:force}, the non-zero eccentricity of our disk automatically decreases the $\varphi$-velocities in the disk relative to the Keplerian velocity. Adjusting for this effect, the $\varphi$-velocities in the disk are 81 percent of the Keplerian values.

As we mention in Section \ref{sec:props}, the internal energy density and mass density floors at $2.27\times 10^{-12}$ may artificially decrease the radial and vertical extent of the disk by providing external pressure support. Therefore, our results at radii within the disk boundaries ($\lesssim 500 R_g$) are more reliable than at larger distances. Without the floors, it is possible that the power law curves for density and pressure in Figures $\ref{fig:radial}$ and $\ref{fig:radial_tilted}$ would continue past $500 R_g$. This additional pressure confinement may also be responsible for the flattening of the disc as compared to the ZEBRA model. Because the floors are non-rotating, the external pressure could decrease the angular momentum of material at the edges of the disk, possibly leading to artificially efficient accretion.

\subsubsection{Bonnerot \& Lu Model}
\label{sec:BonLu}

Recently, \citet{Bonnerot2019} performed a TDE simulation with a realistic stellar trajectory and mass ratio. They found that self-intersections launch outflows. These outflows undergo extensive ``secondary shocks'' that ultimately result in the formation of an accretion disk. This contrasts with our results, in which the formation and circularization of the accretion disk results primarily from stream-disk interactions near pericenter (Section \ref{sec:energy}). Even when violent self-intersections do not occur, as in model TDET30, an accretion disk still forms.

\citet{Bonnerot2019} overcame the numerical challenges of simulating a TDE with a realistic stellar trajectory and mass ratio by using a non-spinning BH and incorporating the local simulation of \citet{Lu2019} into their initial conditions to describe the outflows produced by self-intersection shocks. However, that local simulation includes assumptions that maximize the impact of the self-intersection shocks.

\citet{Lu2019} perform their simulation in a special inertial frame, which they refer to as the simulation box (SB) frame, in which the incoming and outgoing streams collide head-on. The frame is related to the lab frame by a boost to the comoving frame of a local stationary observer at the self-intersection radius followed by a boost to a frame where the $\varphi$-velocity of the outgoing stream vanishes. This technique relies on three assumptions.

First, it requires that the incoming and outgoing streams have equal ($\varphi$-velocities, aspect ratios) or precisely opposite (radial velocities) properties at the point of self-intersection. However, significant pericenter dissipation, Lense-Thirring frame dragging, or hydrodynamic instabilities at the boundary of the stream and the disk could cause the outgoing stream to have a lower density and velocity and a different trajectory relative to the incoming stream, decreasing the violence of the self-intersections.

Second, the incoming and outgoing streams are only completely parallel in the SB frame at the intersection point. As the radial distance from the self-intersection point increases, the head-on gas trajectories used in \citet{Lu2019} diverge from the physical trajectories. Therefore, the approach is only accurate in the case that there are minimal interactions between pre- and post-intersection material and, e.g., the accretion disc. By ignoring these interactions, the head-on approach increases the relative importance of the self-intersections in the overall TDE evolution.

Third, the \citet{Lu2019} simulation uses 2D cylindrical coordinates, implicitly assuming axisymmetry of the colliding streams. This 2D approach also cannot fully capture 3D fluid instabilities and turbulence inherent in the violent interaction.

Many of the novel effects observed by \citet{Bonnerot2019} are tied to the strong outflows sourced at the self-intersection point: for instance, the formation of a retrograde accretion disk with respect to the star's initial orbital angular momentum is due to the preferential loss of prograde debris in the self-intersection outflows. In our simulations, stream-disk interactions near pericenter rapidly become the primary locus of energy dissipation and efficiently suppress the return of coherent outgoing streams to the self-intersection site. This behavior is difficult to reconcile with local mass injection schemes near the self-intersection radius. 

One source of the discrepancy between our work and that of \citet{Bonnerot2019} may be the high-$\beta$ encounter analyzed here, which causes the outgoing stream to significantly expand after the pericenter passage due to differential precession (Section \ref{sec:props}). For a more standard TDE with $\beta \approx 1$, the approach of \citet{Bonnerot2019} may become more accurate. It is also possible that increasing the resolution across the vertical extent of the stream in our simulation would result in stronger nozzle shocks. We leave it to future work to apply our methods to less deeply penetrating TDEs.

\subsection{Analysis of the Tilted TDE Simulation, Model TDET30}
\label{sec:tilted-analysis}

\begin{figure}
	\includegraphics[width=1.0\columnwidth]{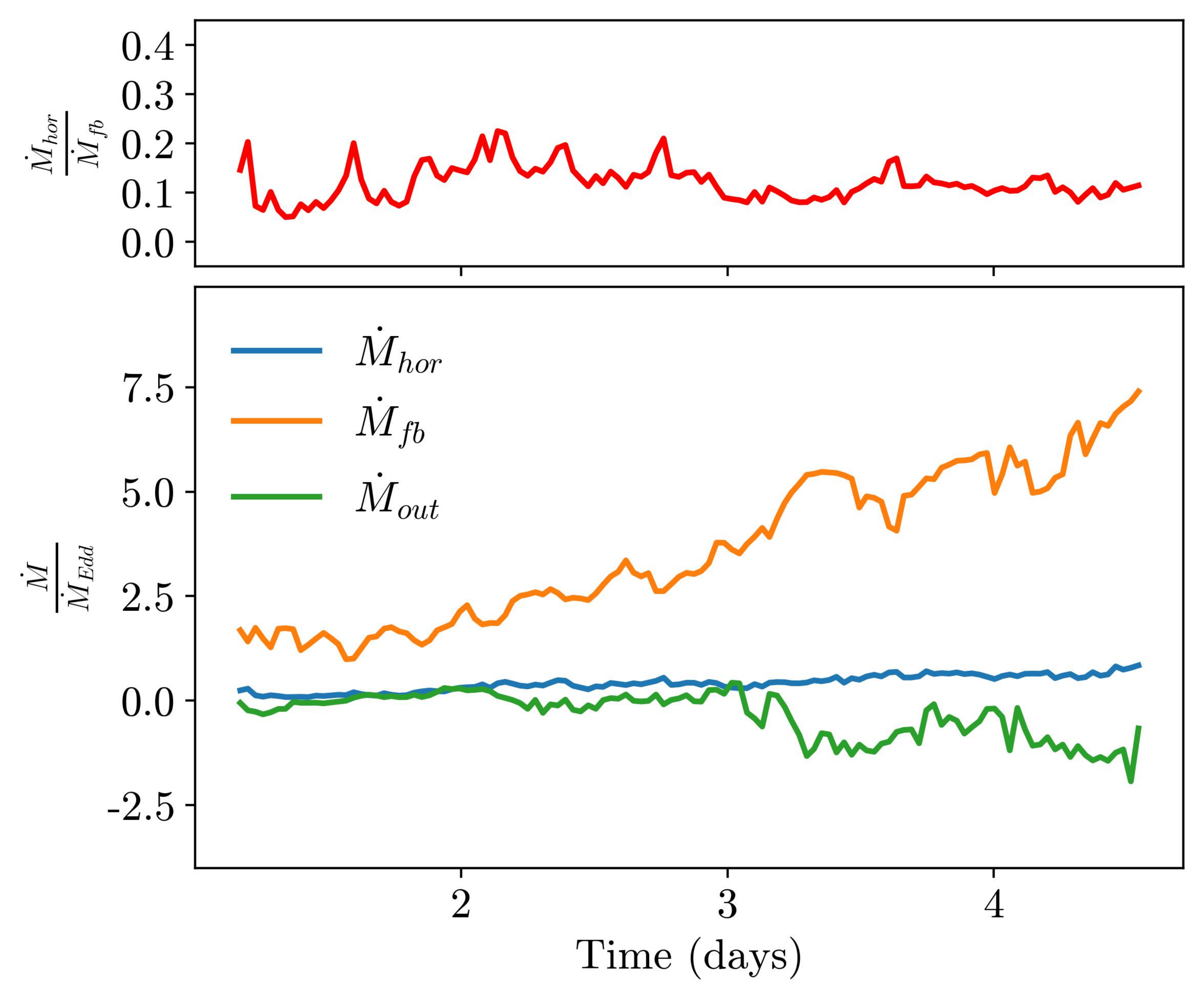}
    \caption{Analogous to Figure \ref{fig:Accretion}, but for the tilted TDE simulation, model TDET30. The accretion efficiency is similar to that of the aligned run.}
    \label{fig:accretion_tilted}
\end{figure}

Figure \ref{fig:accretion_tilted} shows that the accretion efficiency of the tilted TDE is only slightly less than the aligned TDE, hovering from 10 to 15 percent. Just like the aligned scenario, all three mass fluxes increase roughly linearly in time, implying a quadratically increasing disk mass.

\begin{figure*}
	\includegraphics[width=0.8\textwidth]{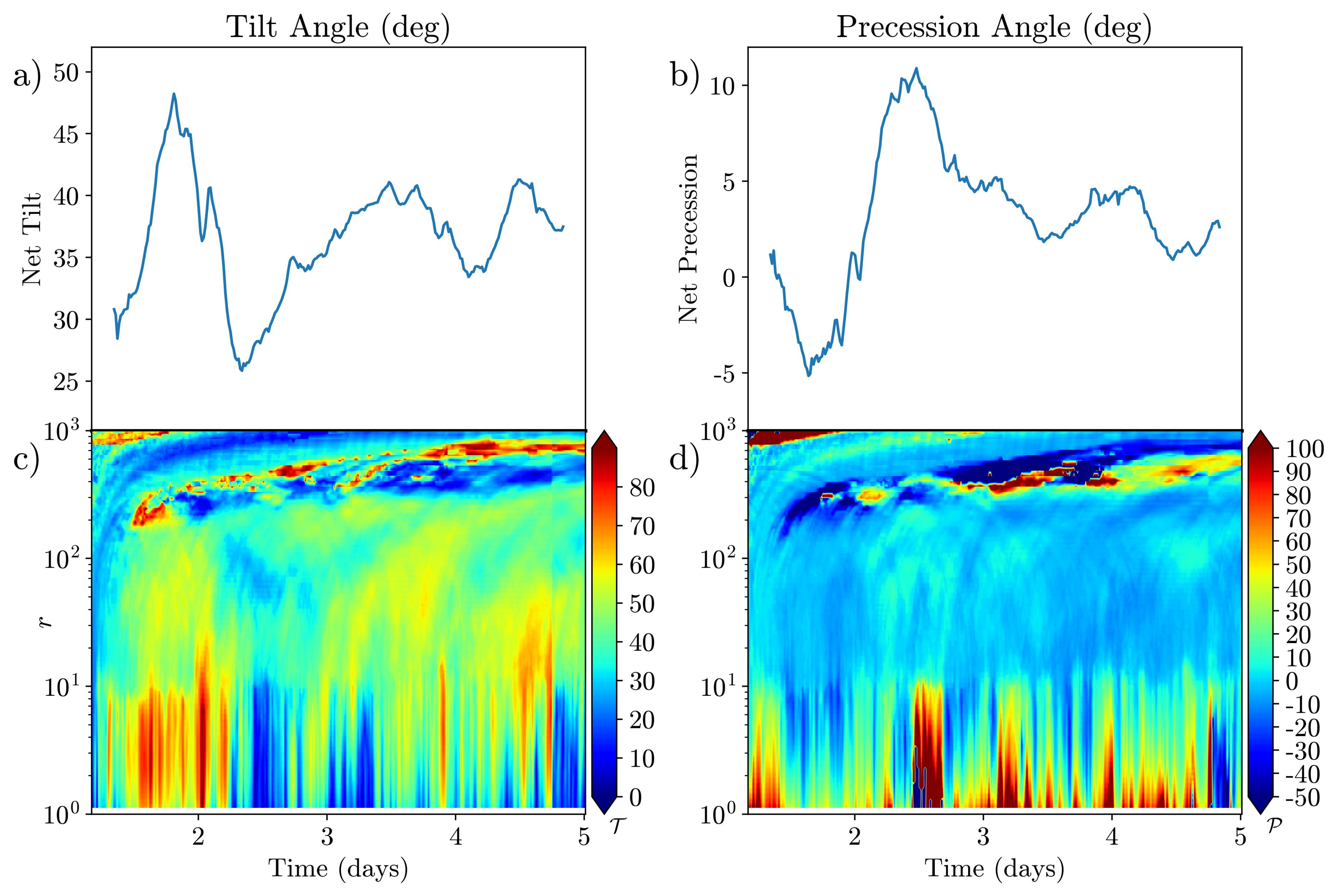}
    \caption{The tilt angle $\mathcal{T}$ and precession angle $\mathcal{P}$ of the tilted disk over time in our titled TDE simulation TDET30. Panels a) and b) depict the tilt and precession angles calculated from the net angular momentum of the disk for material $r<500 R_g$. Panels c) and d) depict the tilt and precession angles calculated from the net angular momentum of the disk material at the radius indicated on the y-axis. Both tilt and precession angles are continuous within the range of radii in the disk outside of stream's pericenter ($10 R_g \lesssim r \lesssim 200 R_g$ at the beginning of the simulation and $10 R_g \lesssim r \lesssim 400 R_g$ at the end of the simulation), indicating that there is no disk-tearing. Below $10 R_g$, tilt and precession angles fluctuate rapidly due to stream-disk interactions at pericenter. We do not observe significant precession of the disk in the duration of the simulation. It is possible that disk precession at small radii is inhibited by the constant injection of stream material in the orbital plane of the star.}
    \label{fig:angles}
\end{figure*}

In Figure \ref{fig:angles}, we compute the tilt and precession angle of the disk using the method of \citet{Fragile2007, Fragile2005, Nelson2000} which we describe below for convenience. The tilt angle $\mathcal{T}$ is given by

\begin{equation}
    \mathcal{T} (r) = \arccos \bigg[ \frac{\mathbf{J}_\textup{BH} \cdot \mathbf{J}_\textup{disk}(r)}{|\mathbf{J}_\textup{BH}||\mathbf{J}_\textup{disk}|} \bigg]
    \label{eq:tilt}
\end{equation}

\noindent
where $\mathbf{J}_\textup{BH} = a M_{\rm BH}\mathbf{\hat{z}}$ is the angular momentum vector of the BH and $\mathbf{J}_\textup{disk}(r)$ is the angular momentum vector of the disk in an asymptotically flat space. $\mathbf{J}_\textup{disk}(r)$ is given component-wise by

\begin{equation}
    (J_\textup{disk})_\rho=\frac{\epsilon_{\mu \nu \sigma \rho} L^{\mu \nu} S^\rho}{2 \sqrt{-S^\alpha S_\alpha}}
    \label{eq:angDsk2}
\end{equation}

\noindent
where

\begin{equation}
    L^{\mu \nu} = \int{\left(x^\mu T^{\nu 0} - x^\nu T^{\mu 0}\right) {\rm d}V}
    \label{eq:L}
\end{equation}

\begin{equation}
    S^\rho=\int{T^{\rho 0} {\rm d} V}
\end{equation}

\noindent
$T$ is the stress-energy tensor, and $\epsilon$ is the 4-dimensional Levi-Civita symbol. The unit vector $\hat{y}$ points along the axis about which the initial conditions are initially tilted and $\hat{z}$ points along the angular momentum axis of the BH. The precession angle is computed similarly using the definition
\begin{equation}
    \mathcal{P}(r) = \arccos \bigg[ \frac{\mathbf{J}_\textup{BH} \times \mathbf{J}_\textup{disk}(r)}{|\mathbf{J}_\textup{BH} \times \mathbf{J}_\textup{disk}|} \cdot \hat{y}  \bigg]
    \label{eq:prec}
\end{equation}

Due to Lense-Thirring precession, we would expect the precession angle of the inner disk to increase monotonically with time. The precession rate of the line of nodes of a particle orbiting with eccentricity $e$ and semi-major axis $a_{\rm orb}$ around a BH with dimensionless spin parameter $a$ is given by \citet{Merritt2013} as
\begin{equation}
    \frac{{\rm d} \Omega_{\rm LT}}{{\rm d} t} = \frac{2G^2M_{\rm BH}^2 a}{c^3 a_{\rm orb}^2 (1 - e)^{3/2}}
\end{equation}
\noindent
Plugging in the dimensionless spin parameter $a=0.9375$ and the mean eccentricity at each radius (Figure \ref{fig:Ecc vs R}), we find that the precession period exceeds the simulation duration of 5 days for radii $r > 85 R_{\rm g}$. The disk in our simulation radially extends to $400 R_{\rm g}$ at late times, so we should only expect significant precession within the inner regions of the disk. However, even at radii $r < 85 R_{\rm g}$, Figure \ref{fig:angles} shows that the precession angle remains consistent throughout the duration of our simulation. One explanation is that precession is inhibited by the angular momentum supplied by the debris stream which is in the initial orbital plane of the star. 

At the times well before peak fallback time considered in our simulation, the fallback rate of the stellar debris increases linearly with time (Figure \ref{fig:accretion_tilted}). Therefore, the debris stream accounts for a significant proportion of the total angular momentum budget. By summing the components of Newtonian angular momentum in the region $r < 500 R_g$, we estimate that the net angular momentum of the stream in this region is approximately 7 percent of the net angular momentum of the disk at late times.

Previous GRMHD simulations of tilted accretion disks have shown disk tearing, where the accretion disk occupies separate planes over different ranges in radii \citep{Liska2019b}. Disk tearing occurs when the torque exerted on the disk by differing rates of Lense-Thirring precession at different radii surpasses the viscous forces holding the disk together. In particular, the inner part of the disk may become aligned with the equatorial plane of the BH, a phenomenon known as Bardeen-Peterson alignment. The continuity of the tilt and precession angles over the range of radii in the disk (Figure \ref{fig:angles}) suggests that the disk remains intact. Similarly to precession, disk tearing and Bardeen-Peterson alignment may be inhibited by the contribution of angular momentum from the stream.

At times well after the end of our simulation, the mass accretion rate will drop below the Eddington limit. Therefore, the disk may cool and begin precessesing and/or tearing. Since this simulation was not run for multiple viscous times of the accretion disk, the presence of precession in tilted TDEs remains an open question.

\section{Conclusions}
\label{sec:conclusions}

In this work, we simulate a tidal disruption of a Sun-like star by a supermassive BH for a realistically large mass ratio ($Q=10^6$) and for a realistic stellar orbit ($e_0 \approx 1$). We simulate the initial disruption in post-Newtonian SPH and migrate to full GRHD as the debris stream approaches the BH. We also present the first simulation of a tilted TDE in GRHD (Section \ref{sec:support}). Our use of realistic parameters poses a number of challenges. A high mass ratio leads to a thin stellar debris stream that is difficult to resolve. We accommodate this difficulty using 2 levels of AMR. A parabolic stellar trajectory necessitates a large range of temporal and spatial scales. As the initial eccentricity of the star increases, the fallback time of the stellar debris and the apocenter of the debris stream orbit grow. The unprecedented efficiency of H-AMR due to GPU-acceleration and AMR allows us to cover the necessary range of scales to simulate the earliest stages of accretion disk formation.

We find that the TDE naturally and efficiently forms an accretion disk, although the high-eccentricity material constantly supplied by the stream inhibits circularization. The accretion efficiency fluctuates between 10 and 20 percent over the duration of our simulation (Figure \ref{fig:Accretion}). We also find that a significant fraction of material accretes at moderate eccentricities ($0.4 < e < 0.7$), with highly eccentric material ($e > 0.7$) preferentially accreted at high latitudes (Figure \ref{fig:ecc-accretion}).

During the post-disruption phase of our aligned TDE simulation, the debris stream undergoes a series of violent self-intersection events in which the incoming and outgoing streams collide. We propose that these self-intersections are the phenomena responsible for the early-time flaring of TDE Swift J1644+57 and other TDEs. The self-intersections account for both the number of flares and their timescale. 

At early times (i.e. during the first 3 days of the simulation), self-intersections play a crucial role in orbital energy dissipation. At late times, the newly formed accretion disk completely intercepts the outgoing stream, causing the violent self-intersection events to cease. As a result, stream-disk interactions near pericenter are the dominant dissipation mechanism. These interactions raise the entropy of the debris stream three quarters of the way to the final entropy of the accretion disk (Figure \ref{fig:diss}).  Consequently, thermal energy flux dominates over mass energy flux in the debris stream post-pericenter compared to before the pericenter passage.

We find that the newborn disk exhibits super-Eddington accretion. The radial and polar dependencies of density and pressure within the disk closely reflect the self-similar solutions proposed by \citet{Coughlin2014} in the ZEBRA model. Non-zero eccentricity has a small effect on the deviation from Keplerian velocities in the disk, and the more prominent effects are from thermal pressure gradient forces (Figure \ref{fig:Force}). The thermal energy generated by accretion heats up the inner part of the disk. The temperature distribution through the disk creates a thermal pressure gradient force that supports the disk against gravity, leading to a sub-Keplerian velocity distribution.

For a TDE with a 30 degree BH-spin--stellar-orbit misalignment angle, we find that nodal precession causes the incoming and outgoing streams to intersect off-center (Figure \ref{fig:splash}). This ejects gas from the outgoing stream onto orbits with larger tilt angles and results in less violent self-intersections. However, an accretion disk still forms with a similar accreted fraction of the material to the aligned case.

The largest drawback of our simulation is its short duration, of about one week, which allows only the small fraction of the stellar debris with specific energy well outside the frozen-in approximation to accrete onto the BH (Section \ref{sec:phantom}). However, these early stages of TDE disk formation are crucial because they capture the initial disk formation and the emergence of the orbital energy dissipation mechanism. Our results suggest that disk formation in TDEs may be a runaway process; once sufficient mass has partially circularized at small radii, stream-disk interactions become the dominant dissipation mechanism, further growing the (initially eccentric) accretion disk. Clearly, the late-time evolution of TDE disks requires further study, particularly as the fallback rate approaches its peak and deviates from linear growth. 

As discussed in Section \ref{sec:params}, it is possible that we under resolve pressure gradients within the stream. This will be corrected in future simulations by including additional levels of AMR. In future work, we plan to study how TDE accretion disk formation is affected by the magnetic field of the disrupted star. This will allow us to model the magnetorotational instability (MRI) of the disk. We also plan to incorporate a variable polytropic index based on the equation of state used by \citet{Shiokawa2015} to more accurately model the thermodynamics of the disk. Finally, we will either use the pseudo-Newtonian potential from \citet{Tejeda2013} in our initial SPH simulation or we will model the entire disruption in full general relativity.

\section{Data Availability}
\label{sec:support}
Movies and 3D renderings are available at MNRAS online and on \href{https://www.youtube.com/playlist?list=PL7YbfRC6zxzAqIPXWbJgXpr4Wq_nwHiDu}{YouTube}. Simulation data is available upon request from Alexander Tchekhovskoy at
\href {mailto:atchekho@northwestern.edu}{atchekho@northwestern.edu}.

\section*{Acknowledgements}
\label{sec:acks}
We thank the anonymous referee for their detailed feedback which has greatly enhanced the clarity of this work. In particular, we thank the referee for the suggestion to analyse the eccentricity of accreting material in more detail. This research was made possible by NSF PRAC awards no. 1615281 and OAC-1811605 as part of the Blue Waters sustained-petascale computing project, which is supported by the National Science Foundation (awards OCI-0725070 and ACI-1238993) and the state of Illinois. Blue Waters is a joint effort of the University of Illinois at Urbana-Champaign and its National Center for Supercomputing Applications. Additional computer time was provided by the Innovative and Novel Computational Impact on Theory and Experiment (INCITE) program under award PHY129. This research used resources of the Oak Ridge Leadership Computing Facility, which is a DOE Office of Science User Facility supported under Contract DE-AC05-00OR22725. 

ERC acknowledges support from NASA through the Hubble Fellowship Program, grant \#HST-HF2-51433.001-A awarded by the Space Telescope Science Institute, which is operated by the Association of Universities for Research in Astronomy, Incorporated, under NASA contract NAS5-26555, and National Science Foundation grant AST-2006684. AT is supported by the National Science Foundation grants AST-1815304 and AST-1911080. NCS received financial support from NASA, through both Einstein Postdoctoral Fellowship Award Number PF5-160145 and the NASA Astrophysics Theory Research Program (Grant NNX17AK43G; PI B. Metzger). He also received support from the Israel Science Foundation (Individual Research Grant 2565/19). ML was supported by John Harvard Distinguished Science Fellowship and ITC Fellowship.

Our simulation, post-processing, and analysis relied on tools from the SciPy and NumPy Python libraries \citep{SciPy,NumPy}. Our 2D and 3D visualizations relied on the Matplotlib Python Library and the VisIt visualization software respectively \citep{Matplotlib, Visit}.

\bibliographystyle{mnras}
\bibliography{main.bib}

\appendix

\section{Derivations}
\label{adx:derivations}

In this appendix, we provide several derivations used in our analysis which did not fit into the main body of the text.

\subsection{Bernoulli Parameter}
\label{sec:bernoulli}

Throughout our analysis, we used the Bernoulli parameter to distinguish between bound and unbound material. The Bernoulli parameter is the ratio of total energy flux to mass energy flux. The total energy flux in spatial coordinate $x^i$ is given by
\begin{equation}
      \Phi_{\textup{total}}^i = -T_t^i = -(\rho c^2 + p + u_g) u_t u^i = -(\rho c^2 + \gamma u_g) u_t u^i
\end{equation}
\noindent
where $T$ is the stress-energy tensor and $g$ is the metric tensor. The total mass energy flux in spatial coordinate $x^i$ is given by
\begin{equation}
  \Phi_{\textup{mass}}^i = \rho c^2 u^i
\end{equation}
\noindent
Therefore, the relativistic Bernoulli parameter is given by
\begin{equation}
b' = \frac{\Phi_{\textup{total}}^i}{\Phi_{\textup{mass}}^i} = -\frac{(\rho c^2 + \gamma u_g) u_t u^i}{\rho c^2 u^i} = -\frac{u_t (\rho c^2 + \gamma u_g)}{\rho c^2}
\end{equation}
Thus defined, the relativistic Bernoulli parameter is counted off from unity, which corresponds to the rest-mass contribution: $b'>1$ corresponds to hydrodynamically unbound and $b'<1$ bound material. To follow the more familiar non-relativistic convention, we subtract the rest-mass contribution so that positive and negative values correspond to unbound and bound material, respectively:
\begin{equation}
    b = b' - 1 = -\frac{u_t (\rho c^2 + \gamma u_g)}{\rho c^2} - 1
\end{equation}

\subsection{Equatorial Geodesics}
\label{adx:geo}

In Figure \ref{fig:dissolve}, we depict a geodesic in the equatorial plane. In the Kerr geometry, equatorial geodesics remain in the equatorial plane, so we set $\theta=\pi/2$ and $u^\theta=0$. We then solve for $u^t$, $u^r$, and $u^\phi$ using the following equations.
\begin{align}
    E= -g_{t\mu}u^\mu
    \label{eq:geo1}\\
    L= g_{\phi\mu}u^\mu
    \label{eq:geo2}\\
    g_{\mu\nu} u^\mu u^\nu = \kappa
    \label{eq:geo3}
\end{align}
\noindent
where $g$ is the metric tensor, $E$ is energy, $L$ is angular momentum and $\kappa=-1$ for time-like geodesics. For the geodesic in Figure \ref{fig:dissolve}, $E$ and $L$ are taken from their simulation values at the Cartesian point $(-500, -200, 0)$. At each point along the geodesic, we compute $u^r$ and $u^\phi$ and integrate the resulting differential equations. We linearlly interpolate the covariant metric to the points along the geodesic.

\subsection{Artificial Velocity Fields}
\label{sec:fields}

In Section \ref{sec:force}, we create artificial velocity fields to control for the effect of the non-zero eccentricity of the disk in our analysis of its velocity distribution. We set up artificial velocity fields of constant eccentricity $e$ and aligned pericenters under a Newtonian regime. For a given point in the midplane, we calculate the semi-major axis of the orbit $a$ and the eccentric anomaly $E$ from the distance from the BH $r$ and the true anomaly $\nu=\theta-\pi/2$.
\begin{align}
    a=\ & \frac{r(1+e\cos{\nu})}{1-e^2}
    \label{eq:axis}\\
    E=\ & \arctan{\frac{\sqrt{1-e^2}\sin{\nu}}{e+\cos{\nu}}}
    \label{eq:ecc_anomaly}
\end{align}
\noindent
Then, we compute the Cartesian state vectors.
\begin{align}
    \mathbf{x}=\ & r \begin{pmatrix}\cos \nu\\ \sin \nu\end{pmatrix}
    \label{eq:pos_cart}\\
    \dot{\mathbf{x}}=\ & \sqrt{\frac{a}{r}} \begin{pmatrix} -\sin E\\ \sqrt{1-e^2} \cos E \end{pmatrix}
    \label{eq:vel_cart}
\end{align}
\noindent
Finally, we compute $\dot{r}$ and $\dot{\varphi}$.
\begin{align}
    \dot{r} =\ & \frac{x_1\dot{x}_1+x_2\dot{x}_2}{r}\\
    \dot{\varphi} =\ & \frac{x_1\dot{x}_2-\dot{x}_1 x_2}{r^2}
\end{align}

\begin{figure}
	\includegraphics[width=1.0\columnwidth]{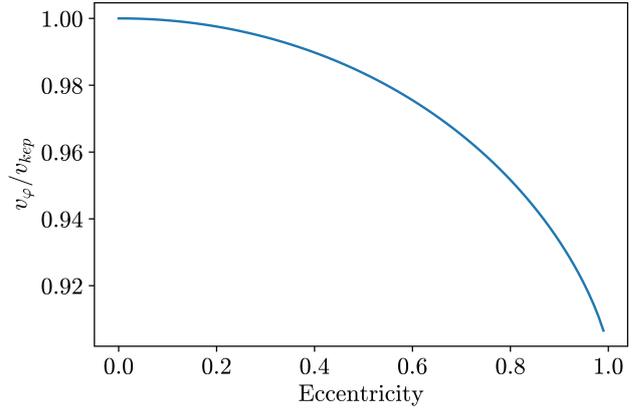}
    \caption{The ratio of $\varphi$-velocity to circular Keplerian velocity for orbits of various eccentricities. We compute $\varphi$-velocity by setting up artificial velocity fields in the equatorial plane with a constant eccentricity. Then, we average $\varphi$-velocity over radius using a mass weight determined by the power law of best fit for the mass density radial profile in the aligned TDE simulation.}
    \label{fig:nonzero}
\end{figure}

\subsection{Analytical Self-Intersection Radius}
\label{sec:Rsi}

The analytical self-intersection radius is given by \citet{Wevers2017}. Consider the orbit of a massless test particle in the equatorial plane around the BH. Averaged over one orbit, general relativistic apsidal precession causes the argument of pericenter to advance by approximately an amount 

\begin{equation}
    \delta \omega = A_S - 2 A_J
\end{equation}

\noindent
where $A_S$ and $A_J$ are the contributions to the apsidal precession of BH mass and spin-induced frame dragging respectively and the precession due to the BH's quadrupole moment is ignored. To the lowest post-Newtonian order, $A_S$ and $A_J$ are given by \citet{Merritt2009} as

\begin{align}
    A_S =\ & \frac{6 \pi}{c^2}\frac{GM_{\textup{BH}}}{r_\textup{p}(1+e)}
    \label{eq:AS}\\
    A_J =\ & \frac{4 \pi a}{c^3}\left(\frac{GM_{\textup{BH}}}{r_\textup{p}(1+e)}\right)^{3/2}
    \label{eq:AJ}
\end{align}
\noindent
where $e$ is the orbital eccentricity. From $\delta \omega$, we find the self-intersection radius with Equation \ref{eq:Rsi}.
\begin{equation}
    R_{\textup{SI}}=\frac{R_\textup{p} (1+e)}{1+e \cos(\pi + \delta \omega / 2)}
    \label{eq:Rsi}
\end{equation}
For a marginally bound stellar orbit $e\approx 1$ with a pericenter radius $r_p = 7 R_{\rm g}$, we find a self-intersection radius of $142 R_{\rm g}$. For a similar orbit with a pericenter radius $r_p = 12 R_{\rm g}$ as seen in our simulation at late times, we find an analytical self-intersection radius of $565 R_{\rm g}$.

\section{Curve Fitting Data}

In this appendix, we provide the curve fitting data for Figures \ref{fig:radial} and \ref{fig:radial_tilted} in Tables \ref{tab:curve_fit} and \ref{tab:curve_fit_tilted} respectively.

\begin{table}
	\centering
	\caption{Curve fitting results for the radial profiles of mass density, pressure, and $\varphi$-velocity from 20-250 $R_g$ shown in Figure \ref{fig:radial}, including the power law parameters ($ax^b$) and their relative standard deviation errors. The exponent for $\varphi$-velocity is fixed at -0.5.}
	\label{tab:curve_fit}
	\begin{tabular}{lcccr}
		\hline
		Variable & a & $\sigma_{\textup{a}}/\mu_{\textup{a}}$ & b & $\sigma_{\textup{b}}/\mu_{\textup{b}}$ \\
		\hline
		$\rho$ & 4.13E-7 & 4.71\% & -1.10 & 1.26\% \\
		pressure  & 2.54E-7 & 7.37\% & -2.32 & 1.00\% \\
		$\varphi$-velocity & 0.759 & 1.64\% & \ & \ \\
		\hline
	\end{tabular}
\end{table}

\begin{table}
	\centering
	\caption{Curve fitting results for the tilted TDE radial profiles of mass density, pressure, and $\varphi$-velocity from 20-250 $R_g$ shown in Figure \ref{fig:radial_tilted}, including the power law parameters ($ax^b$) and their relative standard deviation errors. The exponent for $\varphi$-velocity is fixed at -0.5.}
	\label{tab:curve_fit_tilted}
	\begin{tabular}{lcccr}
		\hline
		Variable & a & $\sigma_{\textup{a}}/\mu_{\textup{a}}$ & b & $\sigma_{\textup{b}}/\mu_{\textup{b}}$ \\
		\hline
		$\rho$ & 5.19E-7 & 1.73\% & -1.34 & 0.397\% \\
		pressure  & 2.91E-7 & 6.75\% & -2.62 & 0.840\% \\
		$\varphi$-velocity & 0.593 & 2.17\% & \ & \ \\
		\hline
	\end{tabular}
\end{table}

\section{Tilting Algorithm}
\label{sec:tilting-algorithm}
\begin{figure}
	\includegraphics[width=1.0\columnwidth]{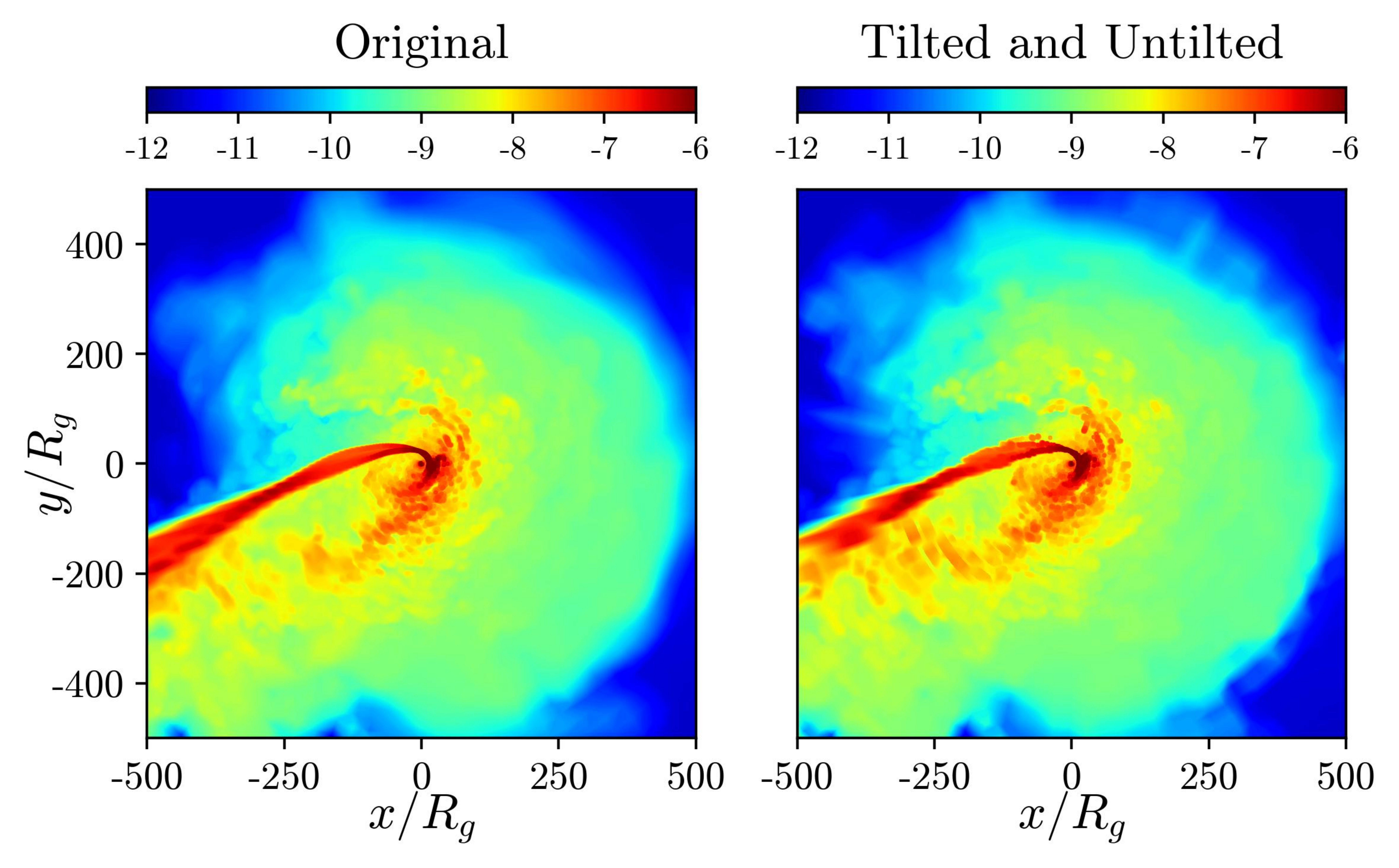}
    \caption{Contour plots of the log of rest mass density in the equatorial plane at 5.7 days. The left panel shows the unaltered data and the right panel shows the data after two applications of the tilting algorithm at angles of $\pi/6$ and $-\pi/6$.}
    \label{fig:tilt_test}
\end{figure}

In this appendix, we describe the tilting algorithm used in our analysis of model TDET30. For our analysis of the tilted TDE, we untilt the data so that the orbital plane of the star lies in the equatorial plane. For each point on our original spherical grid, we convert to Cartesian coordinates and multiply by the rotation matrix $\rm{R}_y(\pi/6)$. Then, we use a third-order spline method to interpolate our data to each point on the rotated grid.

In Figure \ref{fig:tilt_test}, we test our tilting algorithm by tilting and untilting one time slice. While the edges of the stream lose some of their definition, the overall structure of the system remains intact.

\bsp
\label{lastpage}
\end{document}